\documentclass[12pt,preprint]{aastex}

\shortauthors{Matheson et al.}
\shorttitle{GRB~030329 and SN~2003dh---First Two Months}

\begin{document}


\title{Photometry and Spectroscopy of GRB~030329 and Its Associated
Supernova 2003dh: The First Two Months}

\author{T.~Matheson\altaffilmark{1},
P.~M.~Garnavich\altaffilmark{2},
K.~Z.~Stanek\altaffilmark{1},
D.~Bersier\altaffilmark{1},
S.~T.~Holland\altaffilmark{2,3},
K.~Krisciunas\altaffilmark{4,5},
N.~Caldwell\altaffilmark{1},
P.~Berlind\altaffilmark{6},
J.~S.~Bloom\altaffilmark{1},
M.~Bolte\altaffilmark{7},
A.~Z.~Bonanos\altaffilmark{1},
M.~J.~I.~Brown\altaffilmark{8},
W.~R.~Brown\altaffilmark{1},
M.~L.~Calkins\altaffilmark{6},
P.~Challis\altaffilmark{1},
R.~Chornock\altaffilmark{9},
L.~Echevarria\altaffilmark{10},
D.~J.~Eisenstein\altaffilmark{11},
M.~E.~Everett\altaffilmark{12},
A.~V.~Filippenko\altaffilmark{9},
K.~Flint\altaffilmark{13},
R.~J.~Foley\altaffilmark{9},
D.~L.~Freedman\altaffilmark{1},
Mario~Hamuy\altaffilmark{14},
P.~Harding\altaffilmark{15},
N.~P.~Hathi\altaffilmark{10},
M.~Hicken\altaffilmark{1},
C.~Hoopes\altaffilmark{16},
C.~Impey\altaffilmark{11},
B.~T.~Jannuzi\altaffilmark{8},
R.~A.~Jansen\altaffilmark{10},
S.~Jha\altaffilmark{9},
J.~Kaluzny\altaffilmark{17},
S.~Kannappan\altaffilmark{18},
R.~P.~Kirshner\altaffilmark{1},
D.~W.~Latham\altaffilmark{1},
J.~C.~Lee\altaffilmark{11},
D.~C.~Leonard\altaffilmark{19},
W.~Li\altaffilmark{9},
K.~L.~Luhman\altaffilmark{1},
P.~Martini\altaffilmark{14},
H.~Mathis\altaffilmark{8},
J.~Maza\altaffilmark{20},
S.~T.~Megeath\altaffilmark{1},
L.~R.~Miller\altaffilmark{8},
D.~Minniti\altaffilmark{21},
E.~W.~Olszewski\altaffilmark{11},
M.~Papenkova\altaffilmark{9},
M.~M.~Phillips\altaffilmark{4},
B.~Pindor\altaffilmark{22},
D.~D.~Sasselov\altaffilmark{1},
R.~Schild\altaffilmark{1},
H.~Schweiker\altaffilmark{23},
T.~Spahr\altaffilmark{1},
J.~Thomas-Osip\altaffilmark{4},
I.~Thompson\altaffilmark{14},
D.~Weisz\altaffilmark{9},
R.~Windhorst\altaffilmark{10},
and D.~Zaritsky\altaffilmark{11}
}

\altaffiltext{1}{\vspace{-0.15cm}Harvard-Smithsonian Center for Astrophysics, 60
Garden Street, Cambridge, MA 02138}

\altaffiltext{2}{\vspace{-0.15cm}Dept.~of Physics, University of Notre Dame, 225
Nieuwland Science Hall, Notre Dame, IN 46556}

\altaffiltext{3}{\vspace{-0.15cm}Current Address:  Goddard
Space Flight Center, Code 662.20, Greenbelt, MD, 20771-0003}

\altaffiltext{4}{\vspace{-0.15cm}Carnegie Institution of Washington,
Las Campanas Observatory, Casilla 601, La Serena, Chile}

\altaffiltext{5}{\vspace{-0.15cm}Cerro Tololo Inter-American
Observatory, Casilla 603, La Serena, Chile}

\altaffiltext{6}{\vspace{-0.15cm}F.~L.~Whipple Observatory, 670 Mt.~Hopkins Road,
P.O.~Box 97, Amado, AZ 85645}

\altaffiltext{7}{\vspace{-0.15cm}University of California Observatories/Lick
Observatory, University of California, Santa Cruz, Santa Cruz, CA
95064}

\altaffiltext{8}{\vspace{-0.15cm}National Optical Astronomy Observatory, 950 North
Cherry Ave., Tucson, AZ 85719}

\altaffiltext{9}{\vspace{-0.15cm}University of California, Dept.~of
Astronomy, 601 Campbell Hall, Berkeley CA, 94720-3411}

\altaffiltext{10}{\vspace{-0.15cm}Dept.~of Physics and Astronomy, Arizona State
University, Tempe, AZ 85287-1504}

\altaffiltext{11}{\vspace{-0.15cm}Steward Observatory, University of Arizona, 933
N.~Cherry Ave., Tucson, AZ 85718}

\altaffiltext{12}{\vspace{-0.15cm}Planetary Sciences Institute, 620 N.~Sixth Avenue,
Tucson, Arizona 85705}

\altaffiltext{13}{\vspace{-0.15cm}Carnegie Institution of Washington, DTM, 5241 Broad
Branch Road, NW Washington, DC 20015}

\altaffiltext{14}{\vspace{-0.15cm}Carnegie Observatories, 813 Santa Barbara Street,
Pasadena, CA 91101}

\altaffiltext{15}{\vspace{-0.15cm}Dept.~of Astronomy, Case Western Reserve
University, 10900 Euclid Avenue, Cleveland, OH 44106}

\altaffiltext{16}{\vspace{-0.15cm}Dept.~of Physics and Astronomy, Johns Hopkins
University, 3400 N. Charles St., Baltimore, MD 21218}

\altaffiltext{17}{\vspace{-0.15cm}Copernicus Astronomical Center, Bartycka 18,
PL-00-716, Warsaw, Poland}

\altaffiltext{18}{\vspace{-0.15cm}The University of Texas at Austin, McDonald
Obs., 1 University Station C1402, Austin, TX 78712-0259}

\altaffiltext{19}{\vspace{-0.15cm}Five College Astronomy Department,
   University of Massachusetts, Amherst, MA 01003-9305}

\altaffiltext{20}{\vspace{-0.15cm}Universidad de Chile, Casilla 36-D,
Santiago, Chile}

\altaffiltext{21}{\vspace{-0.15cm}Pontificia Universidad Cat{\'o}lica de Chile,
Casilla 306, Santiago, 22, Chile}

\altaffiltext{22}{\vspace{-0.15cm}Princeton University Observatory,
Princeton, NJ 08544}

\altaffiltext{23}{\vspace{-0.15cm}WIYN Consortium Inc., 950 N Cherry
Ave., Tucson, AZ 85719}

\email{\small 
tmatheson@cfa.harvard.edu,
pgarnavi@miranda.phys.nd.edu,
kstanek@cfa.harvard.edu,
dbersier@cfa.harvard.edu,
sholland@milkyway.gsfc.nasa.gov,
kevin@ctiosz.ctio.noao.edu,
ncaldwell@cfa.harvard.edu,
pberlind@cfa.harvard.edu,
jbloom@cfa.harvard.edu,
bolte@ucolick.org,
abonanos@cfa.harvard.edu,
mbrown@noao.edu,
wbrown@cfa.harvard.edu,
mcalkins@cfa.harvard.edu,
pchallis@cfa.harvard.edu,
chornock@astron.berkeley.edu,
Luis.Echevarria@asu.edu,
eisenste@as.arizona.edu,
everett@psi.edu,
alex@astron.berkeley.edu,
flint@dtm.ciw.edu,
rfoley@astron.berkeley.edu,
dfreedman@cfa.harvard.edu,
mhamuy@ociw.edu,
harding@billabong.astr.cwru.edu,
nphathi@asu.edu,
mhicken@cfa.harvard.edu,
choopes@pha.jhu.edu,
impey@as.arizona.edu,
jannuzi@noao.edu,
Rolf.Jansen@asu.edu,
saurabh@astron.berkeley.edu,
jka@camk.edu.pl,
sheila@astro.as.utexas.edu,
rkirshner@cfa.harvard.edu,
dlatham@cfa.harvard.edu,
jlee@as.arizona.edu,
leonard@nova.astro.umass.edu,
weidong@astron.berkeley.edu,
kluhman@cfa.harvard.edu,
martini@ociw.edu,
hmathis@noao.edu,
jose@das.uchile.cl,
tmegeath@cfa.harvard.edu,
mmiller@noao.edu,
dante@astro.puc.cl,
edo@as.arizona.edu,
marina@ugastro.berkeley.edu,
mmp@lco.cl,
pindor@astro.princeton.edu,
dsasselov@cfa.harvard.edu,
rschild@cfa.harvard.edu,
heidis@noao.edu,
tspahr@cfa.harvard.edu,
jet@lco.cl,
ian@ociw.edu,
dweisz@uclink.berkeley.edu,
Rogier.Windhorst@asu.edu,
dzaritsky@as.arizona.edu
}

\begin{abstract}

We present extensive optical and infrared photometry of the afterglow
of gamma-ray burst (GRB) 030329 and its associated supernova
(SN)~2003dh over the first two months after detection (2003 March
30-May 29 UT).  Optical spectroscopy from a variety of telescopes is
shown and, when combined with the photometry, allows an unambiguous
separation between the afterglow and supernova contributions.  The
optical afterglow of the GRB is initially a power-law continuum but
shows significant color variations during the first week that are
unrelated to the presence of a supernova. The early afterglow light
curve also shows deviations from the typical power-law decay.  A
supernova spectrum is first detectable $\sim 7$ days after the burst
and dominates the light after $\sim 11$ days.  The spectral evolution
and the light curve are shown to closely resemble those of SN~1998bw, a
peculiar Type Ic SN associated with GRB~980425, and the time of the
supernova explosion is close to the observed time of the GRB.  It
is now clear that at least some GRBs arise from core-collapse SNe.

\end{abstract}


\section{Introduction}

The mechanism that produces gamma-ray bursts (GRBs) has been the
subject of considerable speculation during the four decades since
their discovery (see M{\'e}sz{\'a}ros 2002 for a recent review of the
theories of GRBs).  The discovery of optical afterglows
(e.g.,~GRB~970228: Groot et al.~1997; van Paradijs et al.~1997) opened
a new window on the field (see, e.g., van Paradijs, Kouveliotou, \&
Wijers 2000).  Subsequent studies of other bursts yielded the
redshifts of several GRBs (e.g.,~GRB~970508: Metzger et al.~1997),
providing definitive evidence for their cosmological origin.
Observations at other wavelengths, especially radio, have revealed
many more details about the bursts (e.g., Berger et al. 2000; Frail et
al. 2003).

Models that invoked supernovae (SNe) to explain GRBs were proposed
from the very beginning (e.g., Colgate 1968; Woosley 1993; Woosley \&
MacFadyen 1999).  There have been tantalizing observational clues that
also pointed to SNe as a possible mechanism for producing GRBs.  The
most direct was GRB~980425: no traditional GRB optical afterglow was
seen, but a supernova, SN~1998bw, was found in the error box of the
GRB (Galama et al. 1998a).  The SN was classified as a Type Ic (Patat
\& Piemonte 1998), but it was unusual, with high expansion velocities
(Patat et al. 2001).  Other SNe with high expansion velocities (and
usually large luminosity as well) such as SN~1997ef and SN~2002ap are
sometimes referred to as ``hypernovae'' (see, e.g., Iwamoto et
al. 1998, 2000).  GRB~980425 was also unusual in the sense that the
isotropic energy of the burst was 10$^{-3}$ to 10$^{-4}$ times weaker
than in classical cosmological GRBs (Woosley, Eastman, \& Schmidt
1999), indicating that this was not a typical burst.

Indirect evidence also relates GRBs to SNe.  Core-collapse SNe are
associated with massive stars (e.g., Van~Dyk, Hamuy, \& Filippenko
1996) and GRBs also appear to be associated with massive stars, based
on their location in their host galaxies (e.g.,~Bloom, Kulkarni, \&
Djorgovski 2002) and statistics of the types of galaxies that host
GRBs (e.g.,~Hogg \& Fruchter 1999).  Chevalier \& Li (2000) have shown
that the afterglow properties of some GRBs are consistent with a shock
moving into a stellar wind formed from a massive star.

The redshift of a typical GRB is $z \approx 1$, implying that a
supernova component underlying an optical afterglow would be
difficult to detect.  At $z \approx 1$, even a bright core-collapse
event would peak at $R > 23$ mag.  Nevertheless, late-time deviations
from the power-law decline typically observed for optical afterglows
have been seen and these bumps in the light curves have been
interpreted as evidence for supernovae (for a recent summary, see
Bloom 2003).  Perhaps the best evidence that classical, long-duration
gamma-ray bursts are generated by core-collapse supernovae was
provided by GRB~011121.  It was at $z = 0.36$, so the supernova
component would have been relatively bright.  A bump in the light
curve was observed both from the ground and with \emph{HST} (Garnavich
et al.~2003a; Bloom et al.~2002).  The color changes in the light
curve of GRB~011121 were also consistent with a supernova (designated
SN~2001ke), but a spectrum obtained by Garnavich et al. (2003a) during
the time that the bump was apparent did not show any features that
could be definitively identified as originating from a supernova.  The
detection of a clear spectroscopic supernova signature was for the
first time reported for the GRB~030329 by Matheson et al. (2003a,
2003b), Garnavich et al. (2003b, 2003c), Chornock et al. (2003), and
Stanek et al.~(2003a).  Hjorth et al. (2003) also presented
spectroscopic data obtained with the VLT.  Their analysis produced
results similar to those presented here.  In addition, Kawabata et
al. (2003) obtained a spectrum of SN~2003dh with the Subaru
telescope.  The properties of the afterglow light curve have also been
described by Burenin et al. (2003), Uemura et al. (2003), and Price et
al. (2003).

The extremely bright GRB~030329 was detected by the French Gamma Ray
Telescope, the Wide Field X-Ray Monitor, and the Soft X-Ray Camera
instruments aboard the \emph{High Energy Transient Explorer II} at
11:37:14.67 (UT is used throughout this paper) on 2003 March~29
(Vanderspek et al.~2003).  With a duration of more than 25 seconds,
GRB~030329 is classified as a long-duration burst (Kouveliotou et
al. 1993).  Peterson \& Price~(2003) and Torii~(2003) reported
discovery of a bright ($R\approx 13$ mag), slowly fading optical
transient (OT), located at $\alpha$~=~$10^{\rm h}44^{\rm m}50\fs0$,
$\delta$~=~$+21\arcdeg31\arcmin17\farcs8$ (J2000.0), and identified
this as the GRB optical afterglow. Due to the brightness of the
afterglow, observations of the optical transient (OT) were extensive,
making it most likely the best-observed afterglow so far.

From the moment the low redshift of 0.1685 for the GRB~030329 was
announced (Greiner et al.~2003), we started organizing a campaign of
spectroscopic and photometric follow-up of the afterglow and later the
possible associated supernova. Stanek et al.~(2003a) reported the
first results of this campaign, namely a clear spectroscopic detection
of a SN~1998bw-like supernova in the early spectra, designated
SN~2003dh (Garnavich et al. 2003c).  In this paper, we report on our
extensive data taken for GRB~030329/SN~2003dh during the first two
months after the burst.

\section{The Photometric Data}

The photometric data are listed in Table \ref{phottable}\footnote{The analysis presented here
supersedes our GCN Circulars by \citet{martini03},
\citet{garnavich03d}, \citet{stanek03}, Li et al. (2003a, b),
\citet{bersier03b}, and \citet{stanek03b}.}.  Much of our
$UBVR_C I_C$ photometry was obtained with the F.~L.~Whipple
Observatory (FLWO) 1.2-m telescope and the ``4Shooter'' CCD mosaic
(Szentgyorgyi et al., in preparation) with four thinned, back-side
illuminated, AR-coated Loral $2048\times2048$ pixel CCDs. The camera
has a pixel scale of $0.335\arcsec$ pixel$^{-1}$ and a field of view of
roughly $11.5\arcmin$ on a side for each chip.  The data were taken in
the $2\times2$ CCD binning mode.  We continuously monitored the
afterglow during the first night in all five bands, obtaining a total
of 149 images.  We also obtained multi-band data each night for the
next 11 nights. We then closely followed the OT in the $R$ band with
only two gaps, when the Moon was very bright or close to the object
and when the ``4Shooter'' was not on the telescope\footnote{All
photometry and spectroscopy presented in this paper are available
through {\tt anonymous ftp} on {\tt cfa-ftp.harvard.edu}, in the
directory {\tt pub/kstanek/GRB030329}, and through the {\tt WWW} at
{\tt http://cfa-www.harvard.edu/cfa/oir/Research/GRB/}.}.

Extensive early $UBVRI$ data were also obtained using an Apogee AP7
CCD camera with the 0.76-m Katzman Automatic Imaging Telescope (KAIT;
Li et al. 2000; Filippenko et al. 2001) at Lick Observatory. The
Apogee camera has a back-illuminated SITe 512$\times$512 pixel CCD
chip, which with a scale of 0$\farcs$8 pixel$^{-1}$ yields a total
field of view of 6$\arcmin.7\times 6\arcmin.7$.  Thirteen $UBVRI$
sets were obtained during the first night, and three sets the next night
(Li et al. 2003a).

Additional $R$-band images, including our earliest photometric data,
were obtained using the Magellan telescopes at Las Campanas
Observatory (LCO) with the LDSS2 imaging spectrograph
\citep{mulchaey01} in its imaging mode, with a scale of $0\farcs378$
pixel$^{-1}$. We also obtained $R$-band data with the LCO Swope 1-m
telescope equipped with the SITe\#3 $2048\times3150$ CCD camera, which
with a scale of 0$\farcs$435 pixel$^{-1}$ yields a total field of view
of 14$\arcmin.8\times 22\arcmin.8$.  Also at LCO, we obtained $BVI$
images with the du~Pont 2.5-m telescope equipped with the TEK\#5
$2048\times 2048$ pixel CCD camera, which with a scale of 0$\farcs$259
pixel$^{-1}$ yields a total field of view of 8$\arcmin.85\times
8\arcmin.85$.

In the $B$ and $R$ bands we obtained a significant number of
images with the KPNO Mayall 4-m telescope equipped with the MOSAIC-1
wide-field camera.  The prime focus Mosaic-1 camera (Muller et
al. 1998) has eight CCDs covering its 36\arcmin$\times$36\arcmin\
field of view. For the majority of the exposures, the telescope was
pointed so that GRB~030329 and photometry reference objects were all
placed on the second of the eight CCDs.  The images were all processed
through the reduction steps listed in version 7.01 of ``The NOAO Deep
Wide-Field Survey MOSAIC Data Reductions'' guide through the
application of a dome flat (Jannuzi et al. in
preparation)\footnote{http://www.noao.edu/noao/noaodeep/ReductionOpt/frames.html}.
The software used for the reductions is described by Valdes (2002).
All of the software is part of the MSCRED software package (v4.7),
which is part of IRAF\footnote{IRAF is distributed by the National
Optical Astronomy Observatory, which is operated by the Association of
Universities for Research in Astronomy, Inc., under cooperative
agreement with the National Science Foundation.}.

Additional late $B$-band data were obtained with the du~Pont 2.5-m
telescope. We also obtained late $B$-band data with the FLWO
1.2-m telescope.

The data were reduced by several of us using different photometry
packages.  We used DoPHOT \citep{sms93}, DAOPHOT II (Stetson, 1987,
1992; Stetson \& Harris 1988), and in some cases the image subtraction
code ISIS (Alard \& Lupton 1998; Alard 2000).  We found excellent
agreement among the various packages.  Images were brought onto a
common zero point using from 10 to $>100$ stars per image, depending
on the filter and depth of the image. We used several field stars
measured by \citet{henden03} to obtain calibrated magnitudes.

In addition, a KAIT calibration of the GRB~030329 field was done on
May 22 UT, 2003 by observing Landolt standard stars \citep{landolt} at
a large range of airmasses under photometric conditions.  Aperture
photometry was performed on these standard star frames in IRAF and
then used to calibrate three local standard stars in the KAIT field of
GRB~030329.  Comparison of the KAIT and the Henden calibrations shows
that they are consistent with each other (to within 0.03 mag).  The
KAIT data were in excellent agreement with the overlapping FLWO data,
with the largest offset of only 0.03 mag in the $V$ band. Such uniform
data allow a great level of detail in analyzing the evolution of the
OT.

In the infrared (IR), the OT was observed with the LCO Swope 1-m
telescope IR camera equipped with Rockwell NICMOS3 HgCdTe
$256\times256$ pixel array with 0$\farcs$6 pixel$^{-1}$ scale,
yielding a 2$\arcmin.5\times 2\arcmin.5$ field of view
\citep{persson95}.  The data were obtained from 2003 April 2 to 10,
using the $J_s$ and $H$ filters.  Typically, three standard stars
\citep{persson98} were observed each night, one each at the beginning,
middle, and end of the night.  We assumed mean values of extinction
appropriate at LCO: $J_s$ (0.10 mag/airmass) and $H$ (0.04
mag/airmass). For a comparison star near the GRB, with brightness
comparable to the OT, this resulted in photometry with a scatter lower
than $0.04\;$mag, indicating accurate and stable photometry for the
whole run.

\section{The Spectroscopic Data}

Spectra of the OT associated with GRB~030329 were obtained over many
nights with the 6.5-m MMT telescope, the 1.5-m Tillinghast telescope at
the F.~L.~Whipple Observatory (FLWO), the Magellan 6.5-m Clay and Baade
telescopes at LCO, the du~Pont 2.5-m telescope at LCO, the Shane 3-m
telescope at Lick Observatory, and the Keck I and II 10-m
telescopes\footnote{The analysis presented here supersedes our GCN and
IAU Circulars by \citet{martini03}, \citet{caldwell03},
\citet{matheson03a}, \citet{garnavich03b}, \citet{matheson03b},
\citet{garnavich03c}, and Chornock et al. (2003).}.  The majority of
the data discussed herein are from the MMT.  The spectrographs used
were the Blue Channel \citep{schmidt89} at MMT, FAST
\citep{fabricant98} at FLWO, LDSS2 \citep{mulchaey01} with Clay, the
Boller \& Chivens \citep{philips02} with Baade, the WFCCD
\citep{weymann99} with du~Pont, the Kast Double Spectrograph
\citep{miller93} at Lick, LRIS \citep{oke95} with Keck I, and ESI
\citep{sheinis02} with Keck II.  Standard CCD processing and spectrum
extraction were accomplished with IRAF.  Except for the April 24
ESI data, all spectra were optimally extracted \citep{horne86}.  The
wavelength scale was established with low-order polynomial fits to
calibration lamp spectra taken near the times of the OT exposures.
Small-scale adjustments derived from night-sky lines in the OT frames
were also applied.  We employed our own routines in IDL to flux
calibrate the spectra; spectrophotometric standards, along with other
observational details, are listed in Table \ref{specjournal}.  We
attempted to remove telluric lines using the well-exposed continua of
the spectrophotometric standards (Wade \& Horne 1988; Matheson et
al.~2000).

The spectra were in general taken at or near the parallactic angle
\citep{filippenko82} and at low airmass (with the obvious exception of
observations from LCO).  The relative fluxes are thus accurate to
$\sim$ 5\% over the entire wavelength range.  The Blue Channel, LDSS2,
and Boller \& Chivens spectrographs suffer from second-order
contamination with the gratings used for these observations.  Through
careful cross-calibration with standard stars of different colors (and
order-sorting filters with the Boller \& Chivens), we believe that we
have minimized the effects of the second-order light.  For the few
nights at the MMT when a broad range of standard stars of different
colors was not available, we used the closest match from either the
preceding or following night.  Comparison with broad-band photometry
indicates that the overall shape of the spectra is correct.

\section{Early Photometry and Spectroscopy: Days 1-12}

The transition between the afterglow and the supernova was gradual, so
we define our ``early'' data based on our observations.  We
obtained spectroscopic data each of the 12 nights between March 30 and
April 10. For each of these nights, we also obtained
multi-band photometric data.

\subsection{Early Photometry}

\begin{figure}
\plotone{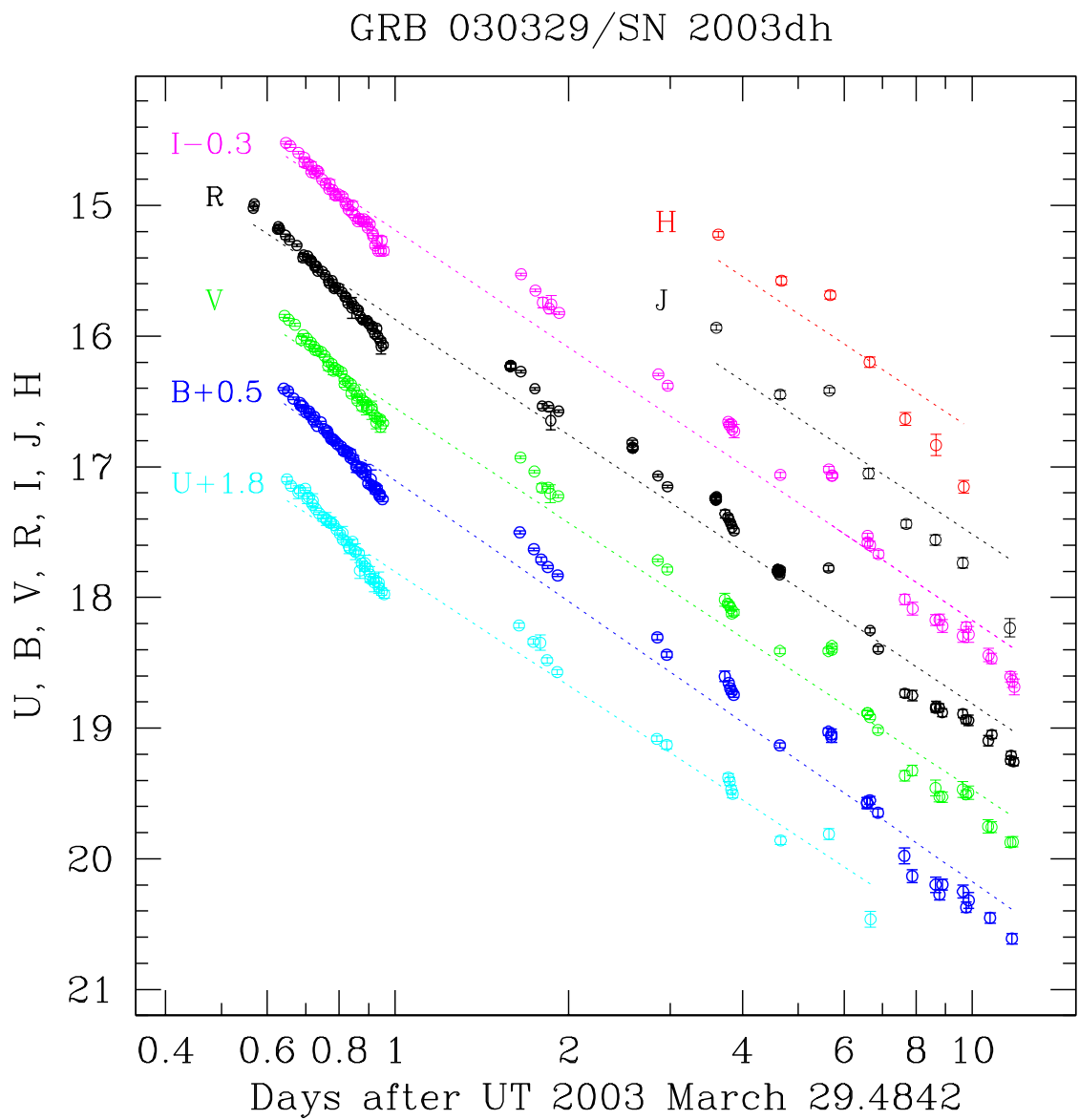}
\caption{Early $UBVRIJH$ light curves of GRB~030329/SN~2003dh (based
on the data in Table \ref{phottable}). The dotted line for each band
is a formal linear fit and is shown only to guide the eye (for $J$ and
$H$, we used the slope of the $R$-band fit). One clearly sees the
bumpy character of the light curve.}
\label{fig_time}
\end{figure}

We plot our GRB~030329 $UBVRIJH$ light curves in Figure
\ref{fig_time}.  Within the first 24 hours, the light curve of the
afterglow consisted of a broken power law typical of many
well-observed bursts (Garnavich et al. 2003d).  But the optical
afterglow exhibited unusual behavior over the following week that has
been analyzed in numerous GCNs (e.g., Li et al. 2003b, c).  As it
is clear that the afterglow cannot be well described with any
semblance of a smooth function usually fitted to describe the OT
evolution, we present and discuss here only our data.  Such uniform
data allow a great level of detail and confidence in analyzing the
evolution of the GRB, including color changes, not usually possible
when using non-homogeneous data compiled from the GCNs and the
literature.

\begin{figure}
\plotone{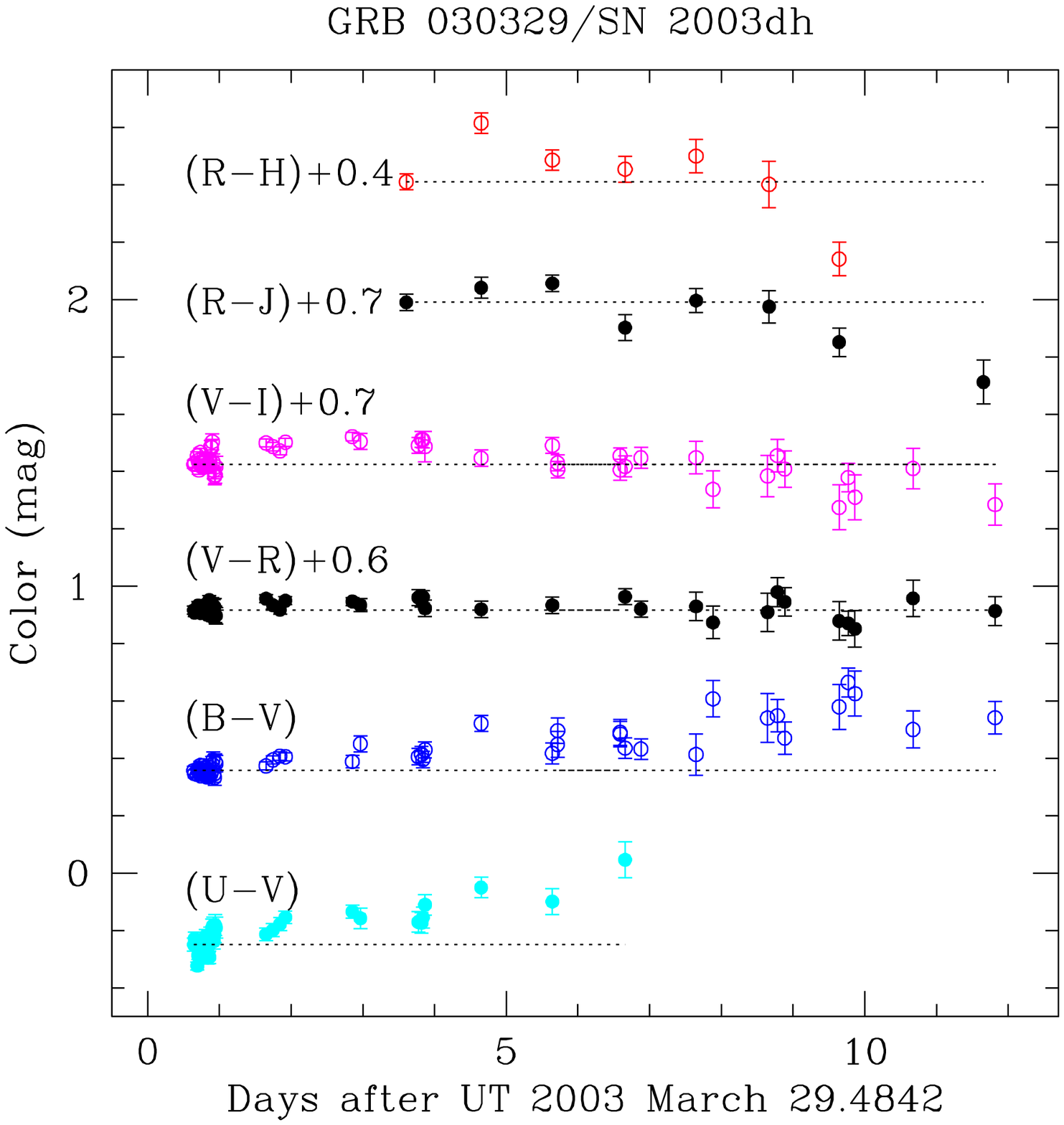}
\caption{Early color evolution of the OT. Fiducial levels
(\emph{dotted lines}) represent the value of the first point for each
color.
\label{fig_color}}
\end{figure}

This is another clear example of an OT changing color as it fades.  A
color change was also seen in the OT of GRB~021004 (Matheson et
al.~2003c; Bersier et al.~2003a).  The color curves of the OT of
GRB~030329 are plotted in Figure \ref{fig_color}, in which the color
changes are more readily apparent (see also Zeh, Klose, \& Greiner
2003).  These changes are discussed in more detail below when we
describe the evolution of the spectral energy distribution (SED).

GRB~030329 is located at Galactic coordinates $l=216\arcdeg\!\!.9867,
b=60\arcdeg\!\!.6997$. To remove the effects of the Galactic
interstellar extinction we used the reddening map of \citet{sfd98}
which yields $E(B-V)=0.025$ mag. This corresponds to expected values of
Galactic extinction ranging from $A_H=0.014$ to $A_U=0.137$ mag, using
the extinction corrections of Cardelli, Clayton, \& Mathis (1989) and
O'Donnell (1994) as prescribed in Schlegel et al. (1998).

\begin{figure}
\plotone{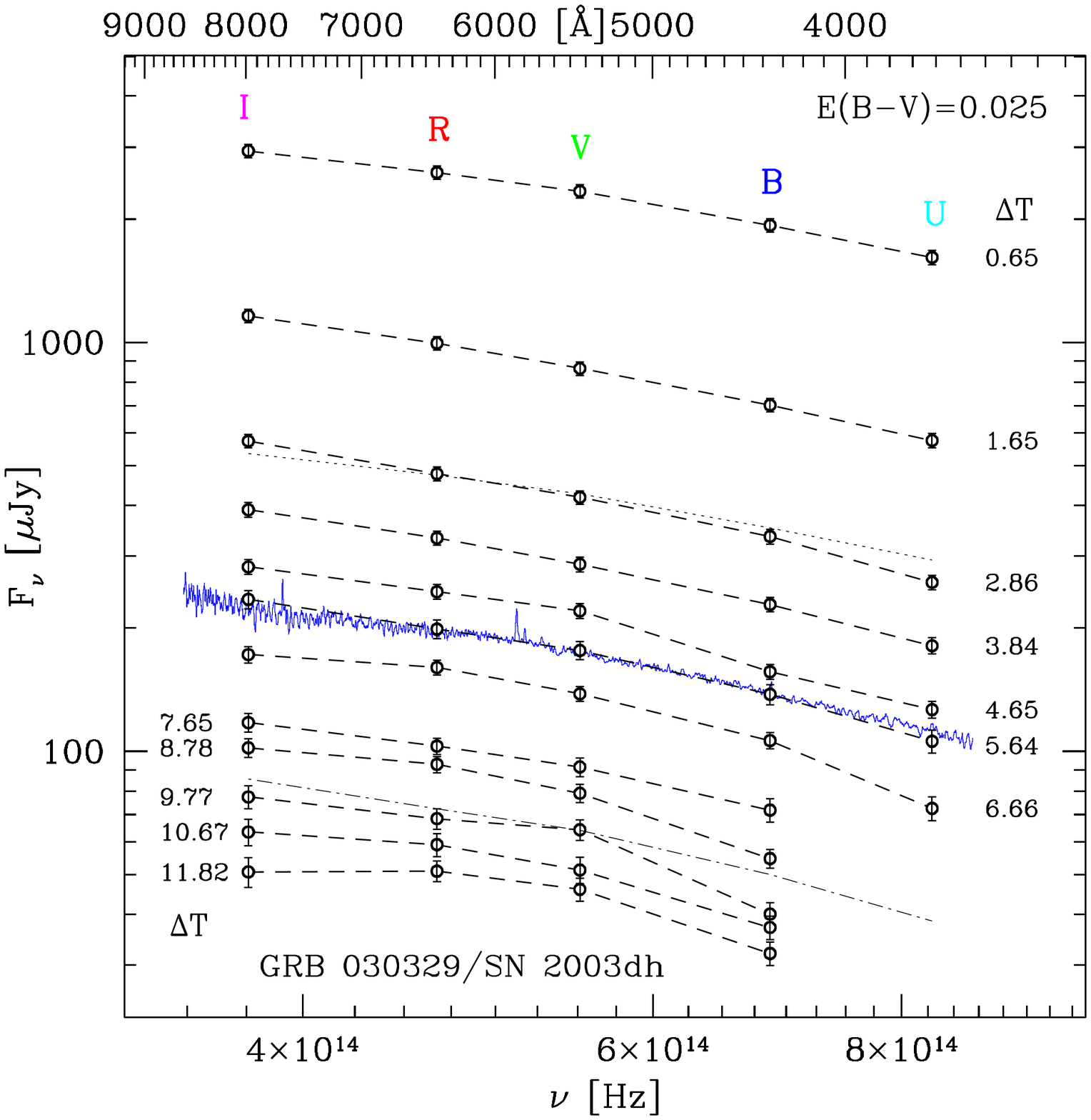}
\caption{Spectral energy distribution (SED) of the optical afterglow
of GRB~030329 at various times (indicated on the right side of each
SED for nights 1-7 and on the left side for nights 8-12).  We
superimposed an MMT spectrum obtained nearly simultaneously with our
photometry at $\Delta T=5.64$ days (our fiducial spectrum). The SED
from $\Delta T=0.65$ days is shown (\emph{dotted line}) on top of the
SED from $\Delta T=2.86$ days. The SED from $\Delta T=5.64$ days is
shown (\emph{dash-dotted line}) on top of the SED from $\Delta T=9.77$
days. For clarity, SEDs from $\Delta T=5.64,9.77,10.67,
\rm{and}~11.82$ days were multiplied by 0.8.
\label{fig_spec}}
\end{figure}

We synthesized the $UBVRI$ spectrum for the first seven nights and
$BVRI$ spectra for later nights from our data by using our best, most
closely spaced measurements for all the nights (Figure \ref{fig_spec}).
We converted the magnitudes to fluxes using the effective frequencies
and normalizations of \citet{fsi95}. These conversions are accurate to
better than 4\%, so to account for the calibration errors we added a
4\% error in quadrature to the statistical error in each flux
measurement.

There are several evolutionary stages to be noticed in Figure
\ref{fig_spec}.  First, the SED gradually evolves between the first
and the third night (see the dotted line in Figure \ref{fig_spec}),
with the spectrum becoming steeper (redder).  The spectral index,
corrected for Galactic reddening of $E(B-V) = 0.025$ mag, changes from
$-0.71$ the first night, through $-0.89$ the second night, to $-0.97$
the third night. Our early shallower slope agrees well with the $-0.66$
slope measured by \citet{burenin03} in their earlier data taken $6-11$
hours after the burst.  Our data are also consistent with the
dereddened spectral slope of $-0.85$ found using SDSS photometry
coinciding with our second night data \citep{lee03}.  Then, at $\Delta
T=4.65$ days (where $\Delta T$ is the time since the GRB), the red
part of the SED ($VRI$) remains unchanged, while the blue part of the
SED ($UB$) is clearly depressed by about $0.1~$mag. On the following
epoch, $\Delta T=5.64$ days, the SED ``recovers'' and resembles
closely the SEDs from nights 3-4.  After $\Delta T=6.66$ days, as
discussed below, the supernova component starts to emerge quickly and
the colors and SEDs undergo dramatic evolution: while nearly unchanged
in $V-R$, the transient becomes more red in $B-V$ and strongly bluer
in $R-I, R-J$, and $R-H$.  Similar color changes at early times
(without $UJH$) were discussed in GCN Circulars by \citet{bersier03b}
and \citet{henden03b}. This peculiar color change is because the
supernova flux peaks around 6000~\AA , raising $V$ and $R$ nearly
equally while the bands redward and blueward slope up toward the
peak.

\begin{figure}
\plotone{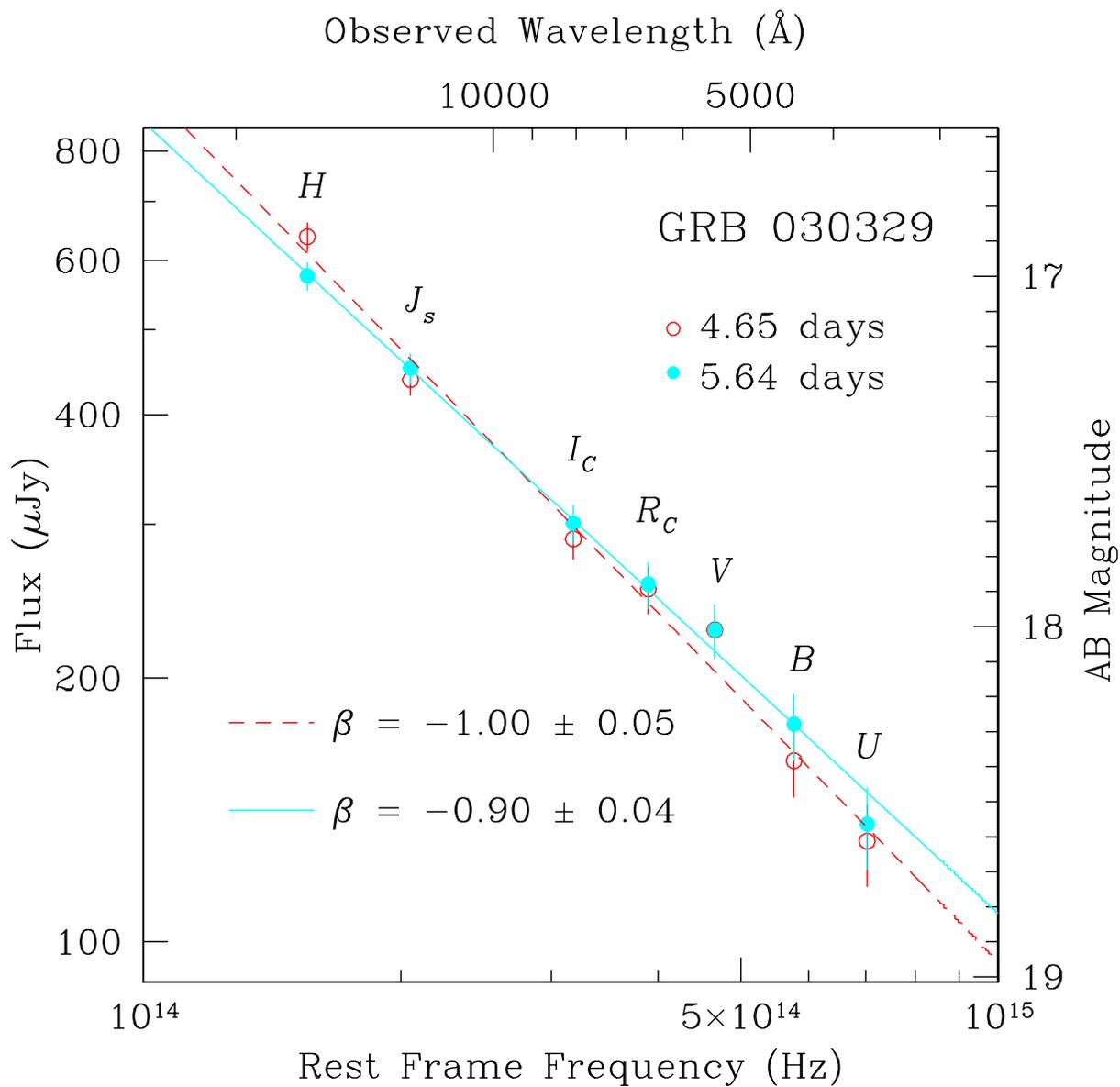}
\caption{Evolution of the SED of the optical afterglow of GRB~030329
between $\Delta T=4.65$ days (\emph{open circles}) and $\Delta T=5.64$
days (\emph{filled circles}), the ``color event'' described in the
text.
\label{fig_ir}}
\end{figure}

The ``color event'' of $\Delta T=4.65$ days is also present in the near-IR
data, as can be seen in Figure \ref{fig_color}.  To highlight this
color change, we show in Figure \ref{fig_ir} the evolution of the SED
of the optical afterglow of GRB~030329 between $\Delta T=4.65$ days
and $\Delta T=5.64$ days.

\subsection{Early Spectroscopy}

\begin{figure}
\epsscale{0.9}
\plotone{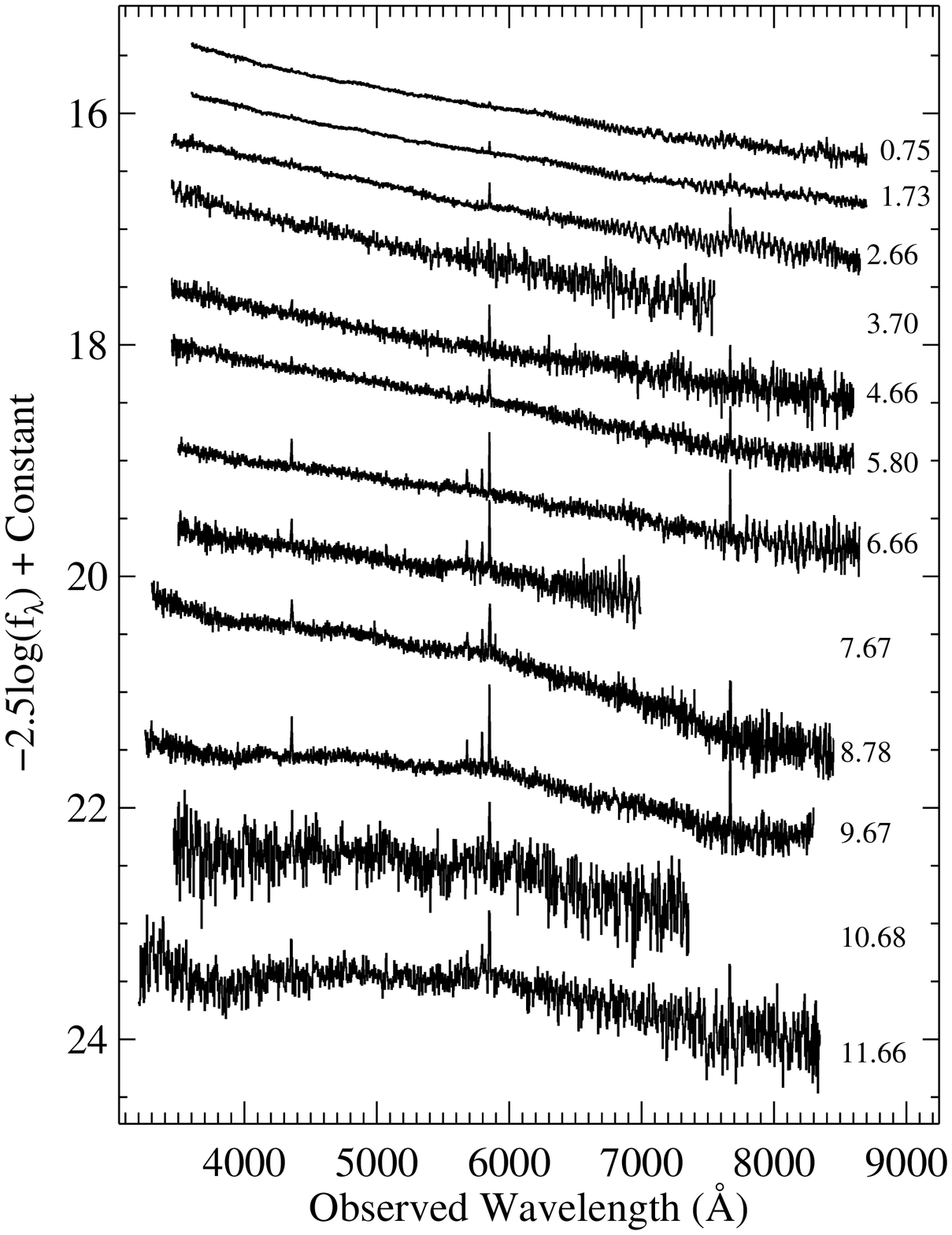}
\caption{Evolution of the GRB~030329/SN~2003dh spectrum, from
March~30.23 UT (0.75 days after the burst), to April 10.14 UT (11.66
days after the burst). The early spectra consist of a power-law
continuum with narrow emission lines originating from \ion{H}{2}
regions in the host galaxy at $z = 0.1685$. Spectra taken after
$\Delta T=6.66$ days show the development of broad peaks characteristic of
a supernova.  In some spectra, regions of bad fringing or low
signal-to-noise ratio have been removed for clarity.  Spectra from
$\Delta T=3.70, 10.68,$ and 11.66 days have been rebinned to improve the
signal-to-noise ratio.  Note that not all spectra listed in Table
\ref{specjournal} are presented in this figure.
\label{grb-all}}
\end{figure}

The brightness of the OT allowed us to observe the OT each of the 12
nights between March 30 and April 10 UT, mostly with the MMT 6.5-m,
but also with the Magellan 6.5-m, Lick Observatory 3-m, LCO du~Pont
2.5-m, and FLWO 1.5-m telescopes.  This provided a unique opportunity
to look for spectroscopic evolution over many nights.  The early
spectra of the OT of GRB~030329 (top of Figure \ref{grb-all}) consist
of a power-law continuum typical of GRB afterglows, with narrow
emission features identifiable as H$\alpha$, [\ion{O}{3}]
$\lambda\lambda$4959, 5007, H$\beta$, and [\ion{O}{2}] $\lambda$3727
at $z = 0.1685$ (Greiner et al.~2003; Caldwell et al.~2003) probably
from \ion{H}{2} regions in the host galaxy.  Assuming a Lambda
cosmology with $H_0 = 70$~km~s$^{-1}$~Mpc$^{-1}$, $\Omega_m$ = 0.3,
and $\Omega_{\Lambda}$ = 0.7, this redshift corresponds to a
luminosity distance of 810 Mpc.

Beginning at $\Delta T=7.67$ days, our spectra deviated from the pure
power-law continuum.  Broad peaks in flux, characteristic of a
supernova, appeared.  The broad bumps are seen at approximately
5000 \AA\ and 4200 \AA\ (rest frame). At that time, the spectrum of
GRB~030329 looked similar to that of the peculiar Type Ic SN~1998bw
a week before maximum light \citep{patat01} superposed on a
typical afterglow continuum. Over the next few days the SN features
became more prominent as the afterglow faded and the SN brightened
toward maximum.

\section{Later Photometry and Spectroscopy: Days 13-61}

\subsection{Later Photometry}

\begin{figure}
\plotone{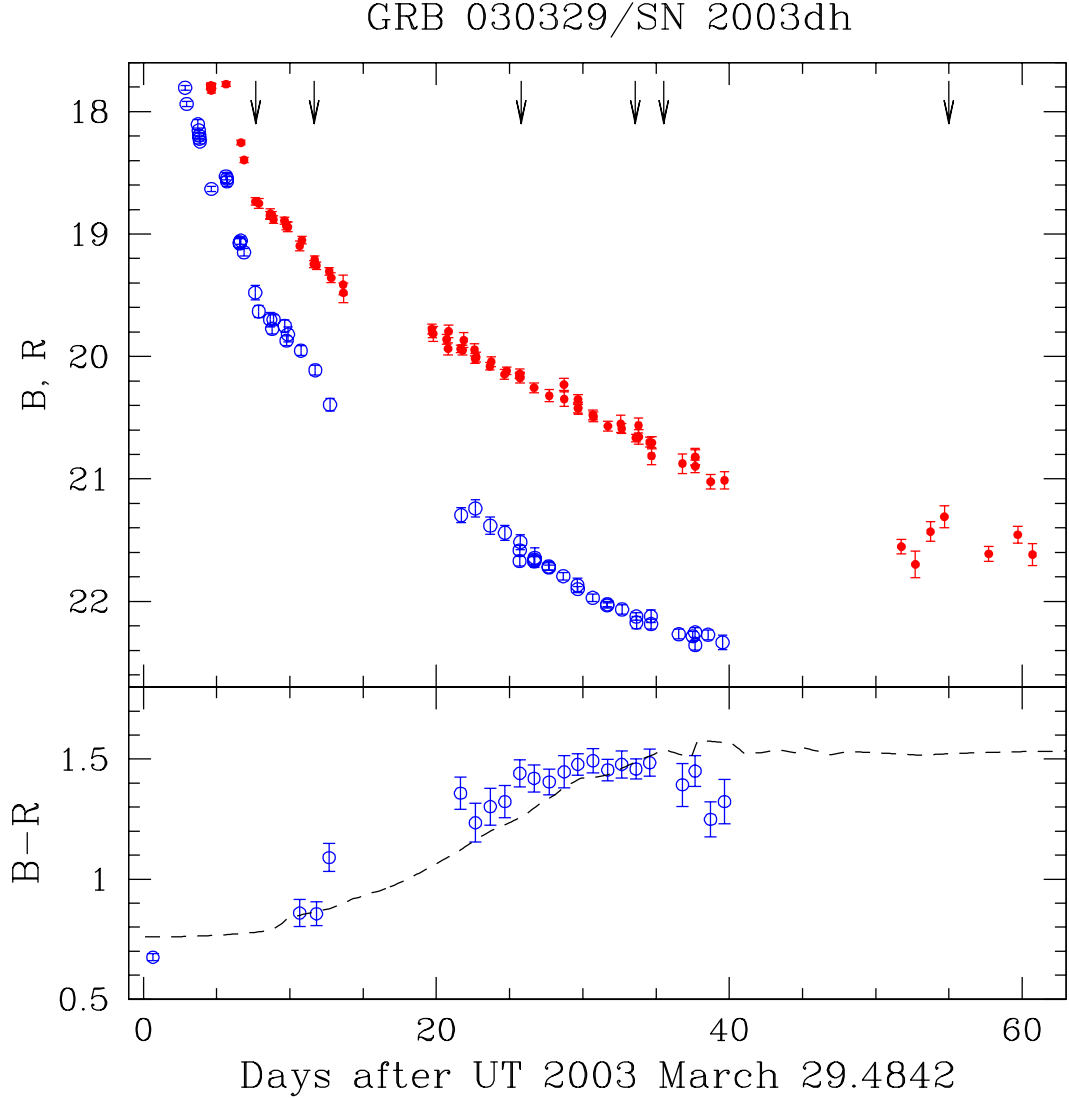}
\caption{Upper panel: $B$-band (\emph{open circles}) and $R$-band
(\emph{filled circles}) photometry at later times, with the last
$R$-band epoch at $\Delta T=60.7$ days after the burst. The first two
arrows correspond to the first time when the supernova signature could
be seen in the spectra and the last spectrum in the continuous
series. The remaining arrows correspond to our spectra taken after
$\Delta T=12.0$ days. Lower panel: $B-R$ color evolution in later
times.  The solid line indicates the color expected for an afterglow
with a fixed power-law spectrum plus a supernova like SN~1998bw
K-corrected to the redshift of GRB~030329. Contamination from the host
galaxy may contribute to the $B$-band light at late times.
\label{later}}
\end{figure}

We continued observing the OT in the $R$ band using mostly the FLWO
1.2-m telescope, and also obtaining some data with the KPNO 4-m and
the LCO Swope 1-m telescopes. In the $B$ band, we obtained most
of the later data with the KPNO 4-m, and also some data with the
du~Pont 2.5-m and the FLWO 1.2-m telescopes. The two gaps in the
$R$-band coverage correspond to the Moon being bright or near the
position of the OT, and also when the CCD camera was not mounted. The
results are shown in Figure \ref{later}.

There are several interesting features to be seen in
Figure \ref{later}. Coinciding with the first detection of the supernova
in the spectra, both $R$ and $B$ light curves start to decay more slowly
(this can be also seen in Figure \ref{fig_time}).  In addition,
the $(B-R)$ color undergoes a dramatic change at later times, as can be
seen in the lower panel of Figure \ref{later}. Both of these
characteristics result from the supernova component, redder in $(B-R)$
color than the GRB afterglow, strongly contributing to the total light
of the OT starting at $\Delta T=7.67$ days. The strong $(B-R)$ color change
indicates that at later times the supernova component dominates the
total light, as will be discussed in more detail later in the paper.

Another striking feature is the ``Jitter Episode'' (Stanek, Latham, \&
Everett 2003; Stanek et al. 2003c; Ibrahimov et al. 2003) 
in the late $R$-band light curve
observed between 51.75 and $60.7\;$days after the burst.  The light
curve is seen to vary on timescales of $\sim 2\;$days by $>0.3\;$mag,
such as when the OT brightens from $R=21.70 \pm 0.11$ mag at $\Delta
T=52.71$ days to $R=21.31 \pm 0.09$ mag at $\Delta T=54.69$ days, only
to fade to $R=21.61 \pm 0.06$ mag at $\Delta T=57.72$ days.

We should stress that these data were obtained with exactly the same
instrumentation and reduced with the same software and in the same
manner as our earlier, much smoother data (see Figure \ref{later}).
This ``Jitter Episode'' is unusual when compared to the whole data
set and we strongly believe that it is real. We will discuss it in
more detail later in the paper.

\subsection{Later Spectroscopy}

\begin{figure}
\epsscale{0.9}
\plotone{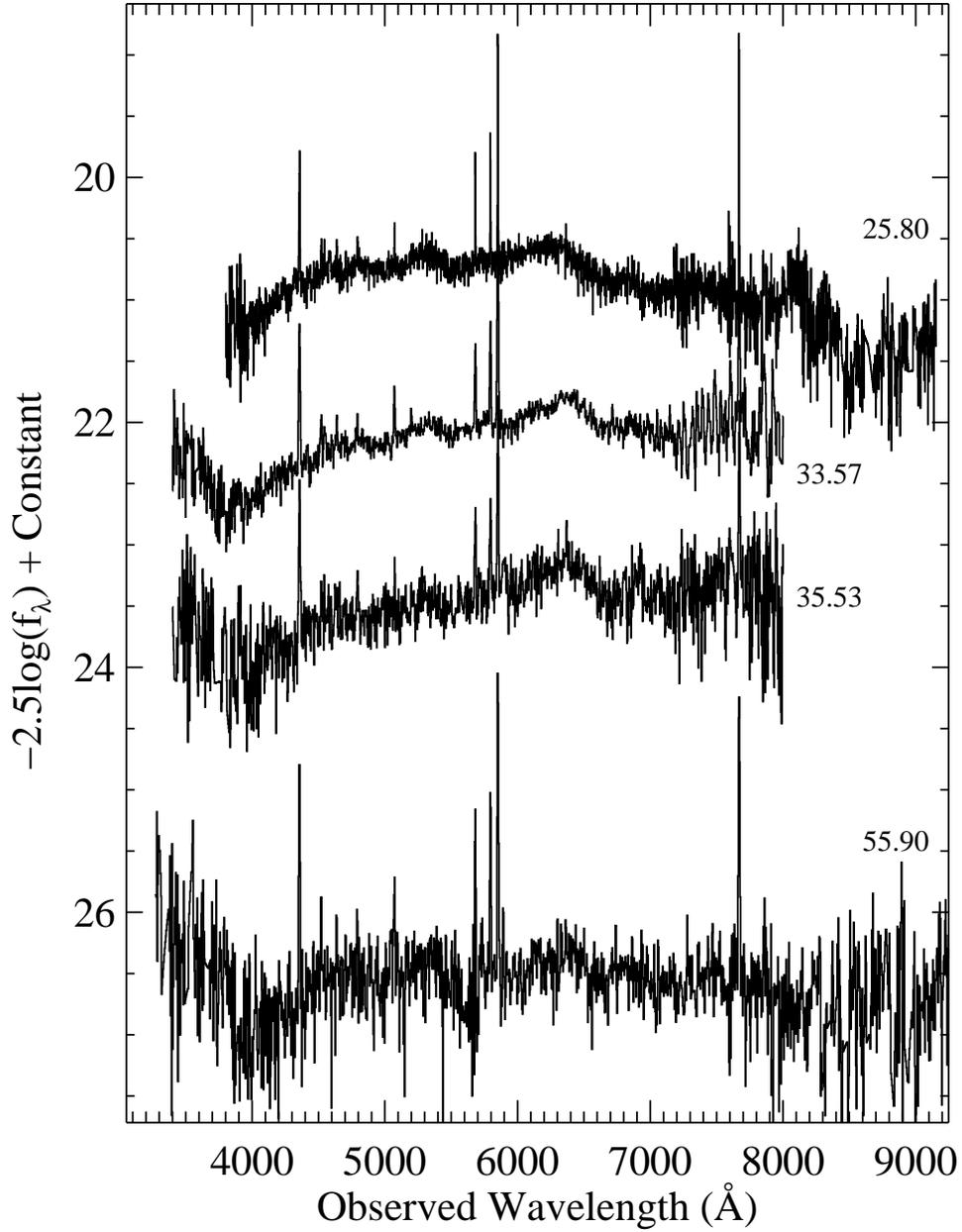}
\caption{Evolution of the GRB~030329/SN~2003dh spectrum, from April
24.28 UT (25.8 days after the burst), to May 24.38 (55.9 days after
the burst).  The power-law contribution decreases and the spectra
become more red as the SN component begins to dominate, although the
upturn at blue wavelengths may still be the power law.  The broad
features of a supernova are readily apparent, and the overall spectrum
continues to resemble that of SN~1998bw several days after maximum.
The $\Delta T=25.8$ days spectrum is a combination of the $\Delta
T=25.71$ days MMT spectrum and the $\Delta T=25.89$ days ESI spectrum.
The dip near 5600 \AA\ in the $\Delta T=55.90$ days spectrum is due to
the dichroic used in LRIS, and is not intrinsic to the OT.
\label{spectra-later}}
\end{figure}

Later spectra obtained on April 24.28, May 2.05, May 4.01, and May
24.38 continue to show the characteristics of a supernova.  As
the power-law continuum of the GRB afterglow fades, the supernova
spectrum rises, becoming the dominant component of the overall
spectrum (Figure \ref{spectra-later}).

\section{Analysis}

\subsection{Properties of the Host}

The low redshift of this burst meant that the rest-frame
optical spectrum of the host galaxy could be obtained, thus allowing
the use of well-tested techniques for measuring the metallicity,
reddening, and star-formation rate of the host.  The MMT spectra from
the nights of 2003 Apr 4, 5, 7, and 8 UT were averaged together and a
low-order fit to the continuum was subtracted, since the SN was
apparent in the averaged spectrum.  At a later date when the optical
transient has completely faded, it will be valuable to get a spectrum
showing the absorption-line component, but the present spectrum is
suitable for studying the emission-line component.  \emph{HST} images and
spectra (Fruchter et al. 2003) show the host to extend to
about 0.5\arcsec, so most of the light of the host galaxy should be
contained within the MMT slit used even though the GRB was off center.
The H$\alpha $ flux measured should thus refer to the entire
galaxy, at least for those nights where the seeing was good.

Overall, the emission-line spectrum shows strong forbidden oxygen
lines and hydrogen Balmer lines, but no detection of the [\ion{N}{2}]
$\lambda \lambda$6548, 6584 lines, with line ratios indicative of
low-metallicity gas photoionized by stars.  Table \ref{ratiotable}
lists the observed ratios, measured using the IRAF ``splot'' routine.
We first estimate the reddening using the H$\alpha $/H$\beta $
line-intensity ratio as a measure of the Balmer decrement, assuming
Case B recombination (e.g., Osterbrock 1989) and the Whitford
extinction law.  The redshifted H$\alpha $ line is affected by
telluric absorption, which may have lead to errors in the
H$\alpha$/H$\beta $ ratio, but the dominant source of error in the
line ratios is simply photon counting.  The reddening implied by the
difference between the observed ratio and the theoretical value is in
the range $E(B-V)$ = 0.05 - 0.11 mag. If the Galactic foreground
reddening is $E(B-V)$ = 0.025 mag (Schlegel et al. 1998), then the
reddening intrinsic to the host is in the range 0.03 to 0.09 mag.

To estimate the oxygen abundance in the host, we use the $R_{23}$ method
that employs the ratios of [\ion{O}{3}]/[\ion{O}{2}],
[\ion{N}{2}]/[\ion{O}{2}], and [\ion{O}{2}]+[\ion{O}{3}]/H$ \beta $ (Pagel
1986; Kewley \& Dopita 2002).  Using a reddening of 0.05 mag to
correct the line ratios, and the parameterizations found in Kewley \&
Dopita, we derive an ionization parameter $q = 2 \times 10^7$ cm
s$^{-1}$, which then leads to an oxygen abundance of log(O/H)+12 =
8.5, or about 0.5 Z$_{\sun} $.  The lack of detectable [\ion{N}{2}] is
consistent with this moderate metallicity.

The H$\alpha$ flux was measured from the spectrum of 2003 April 8, and
when corrected for reddening, gives an H$\alpha$ luminosity of
$L$(H$\alpha) = 6.6\times 10^{40}$ erg s$^{-1}$, for a distance of 810
Mpc.  This corresponds to a current star formation rate of $7.9\times
10^{-42}$ $L$(H$\alpha$) = 0.5 M$_{\sun} $ yr$^{-1}$ (Kennicutt 1998).
This would be a modest star formation rate in a large galaxy, but the
host of GRB~030329 is probably a dwarf. Fruchter et al. (2003)
estimate the magnitude of the host to be $V = 22.7$, meaning $M_V =
-16.9$ mag, which is similar to the luminosity of the SMC.  The
moderate metallicity we have calculated is in accord with the galaxy
luminosity, in this case corresponding well with that of the LMC,
which has a metallicity of log(O/H)+12 = 8.4 (Russell \& Dopita 1990).

Star formation rates in dwarfs can vary widely.  Hunter, Hawley, \&
Gallagher (1993) report a range of rates from 0.001 to 3 M$_{\sun} $
yr$^{-1}$, the latter limit referring to starburst dwarfs such as
NGC~1569.  The mean value is around 0.03 M$_{\sun} $ yr$^{-1}$.  A
useful comparison of star formation ability in local dwarf galaxies
can then be made by calculating the birthrate, the ratio of the
current star formation rate to the average past rate.  We estimate
this quantity simply by normalizing the current rate to the galaxy
blue luminosity divided by an age of 12 Gyr, and assuming $M/L =
3$. For the host of GRB~030329, the birthrate is about 5 M$_{\sun} $
yr$^{-1}$.  That can be compared to the SMC value of 0.3 M$_{\sun} $
yr$^{-1}$, and the value of 2 M$_{\sun} $ yr$^{-1}$ for the starburst
galaxy NGC~1569 (derived from data in Hunter et al. 1993 and Kennicutt
\& Hodge 1986).  One is driven to the conclusion that the GRB~030329
host is also a starburst dwarf galaxy.

The more massive hosts of other GRBs also show large star formation
rates, particularly when measured via radio or sub-mm
techniques. Berger et al. (2003) calculate rates of 100-500 M$_{\sun}
$ yr$^{-1}$ in bolometrically luminous hosts ($L > 10^{12}$ L$_{\sun}
$). However, the rates derived from an optical emission line
([\ion{O}{2}] $\lambda 3727$) for other GRB hosts are much
lower, $1-10$ M$_{\sun} $ yr$^{-1}$ (Djorgovski et al. 2001).  The
discrepancy in rates is not yet understood, particularly since the
extinction measured in the optical for the Djorgovski hosts is low, as
we have found here.  Sub-mm observations of the host of GRB~030329
would be interesting in this regard.

\subsection{Extinction Toward the GRB in the Host}

We used our $UBVR_CI_CJ_sH$ photometry from $\Delta T = 5.64$ days
after the burst to investigate whether there is any evidence for
extinction in the host galaxy along the line of sight to
\objectname{GRB~030329}/\objectname{SN~2003dh}.  The optical and
infrared magnitudes were converted to flux densities based on the AB
corrections given in \citet{fsi95} and \citet{M1995}.  Each data point
was corrected for a small Galactic reddening of $E(B-V) = 0.025 \pm
0.020$ mag (Schlegel et al. 1998).  No corrections were applied for
any reddening that may be present in the host galaxy or in
intergalactic space between us and the host.

The spectral energy distribution was fit by $f_{\nu}(\nu) \propto
\nu^{\beta} \times 10^{-0.4 A(\nu)}$, where $f_\nu(\nu)$ is the flux
density at frequency $\nu$, $\beta$ is the intrinsic spectral index,
and $A(\nu)$ is the extragalactic extinction along the line of sight
to the burst.  The dependence of $A(\nu)$ on $\nu$ has been
parameterized in terms of the rest-frame $A_B$ following the three
extinction laws given by \citet{P1992} for the Milky Way (MW), the
Large Magellanic Cloud (LMC), and the Small Magellanic Cloud (SMC).
The fit provides $\beta$ and $A_B$ simultaneously for each of the
assumed extinction laws.  The unextinguished case ($A_B=0$) was also
considered.

\begin{figure}
\plotone{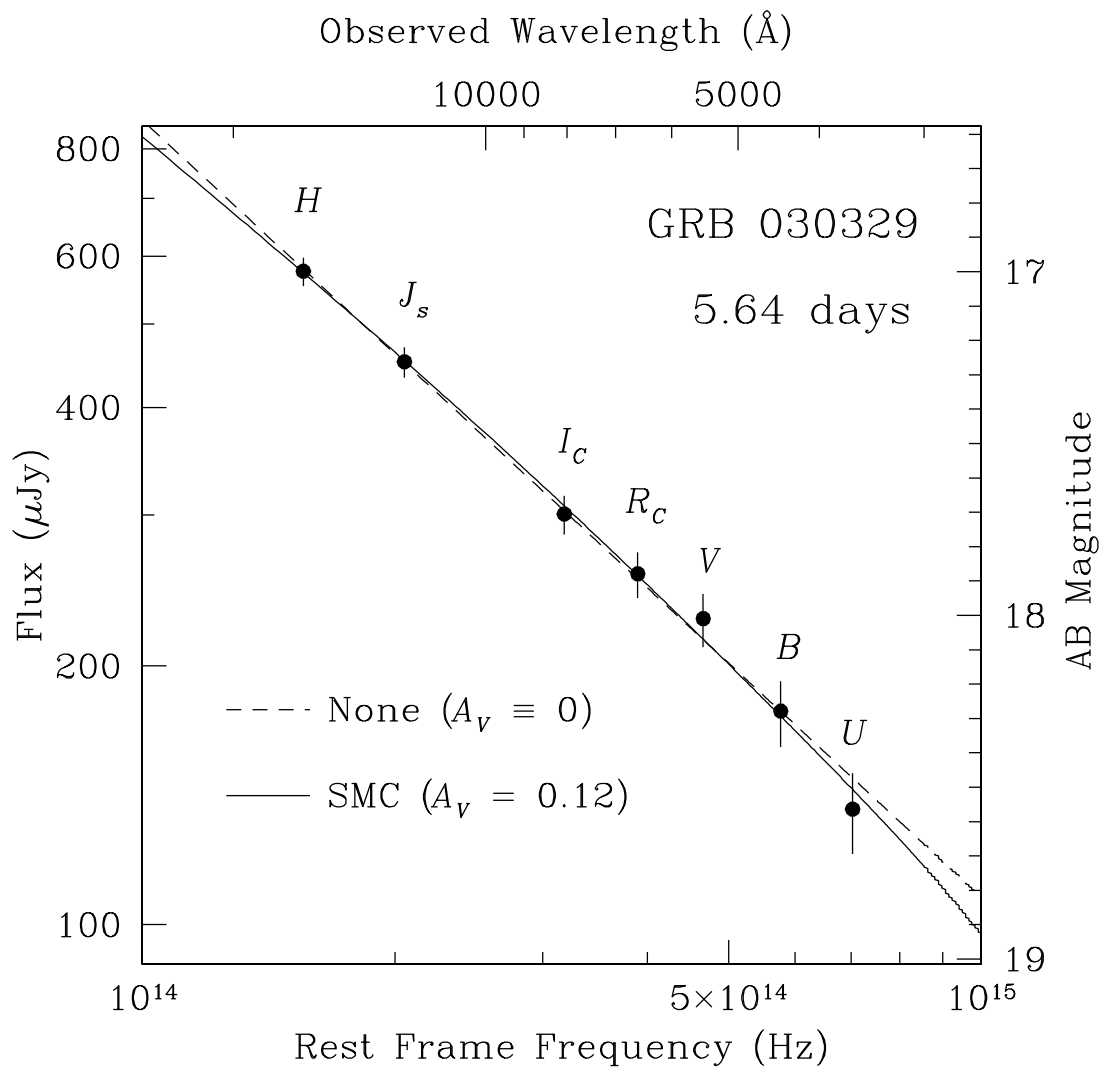} \figcaption[./sed.eps]{Spectral energy distribution
of the optical afterglow of \protect\objectname{GRB
030329}/\protect\objectname{SN 2003dh} at $\Delta T = 5.64$ days after
the burst.  The filled circles represent observed photometry corrected
for extinction in the Milky Way and shifted to the rest frame of the
host galaxy.  The lines represent the best-fitting spectral energy
distribution, assuming an SMC extinction law (\emph{solid line}) or no
extinction (\emph{dashed line}).  If we assume that the unextinguished
spectrum follows $f_\nu(\nu) \propto \nu^{\beta}$ then the best fit
has $\beta_0 = -0.80 \pm 0.20$ and $A_B = 0.16 \pm 0.30$
mag.\label{FIGURE:sed}}
\end{figure}

The best fit is for an SMC extinction law with $A_B = 0.16 \pm
0.30$ mag of extinction in the host and an intrinsic spectral slope of
$\beta = -0.80 \pm 0.20$ ($\chi^2/\mathrm{DOF} = 0.267$).  All three
extinction laws of \citet{P1992} give fits that are statistically
similar($\chi^2/\mathrm{DOF} = 0.267$--0.282) and consistent with $A_B
= 0.16$ mag and $\beta = -0.80$.  Therefore we are unable to constrain the
form of the extinction law in the host.  This slope is also consistent
with the no extinction case ($A_B = 0$ with $\chi^2/\mathrm{DOF} =
0.273$).  Therefore, we conclude that there is no strong evidence for
extragalactic dust along the line of sight between us and
\objectname{GRB 030329}.  Figure \ref{FIGURE:sed} shows the SED at 5.64
days along with fits for an SMC extinction law and no extinction.  To
test for dust along the line of sight between us and the host we
repeated our fits allowing the redshift of the dust to be a free
parameter.  The best fit was for $z = 0.00 \pm 0.09$ with $A_B = 0.17
\pm 0.31$ mag and $\beta = -0.81 \pm 0.18$ ($\chi^2/\mathrm{DOF} = 0.352$).

The most likely distribution for the dust is an SMC extinction
law with $A_B = 0.16 \pm 0.30$ mag in the host galaxy, which
corresponds to $A_V = 0.12 \pm 0.22$ mag and $E_{B\!-\!V} = 0.04 \pm
0.08$ mag in the rest frame of the host.  Note that this is consistent
with the reddening derived from line ratios in the previous section.


\subsection{Evidence for a Cooling Break}

\citet{pfk2003} find that the slope of the optical decay increases
from $\alpha = -0.87 \pm 0.03$ to $\alpha = 1.97 \pm 0.12$
approximately 0.5 days after the burst.  If their interpretation of
this as evidence for a jet break is correct, then the expected
electron index is $p \approx 2$ since $\alpha = p$ after the jet break
has occurred.  \citet{tmg2003} report that \emph{Rossi-XTE\/} $X$-ray
observations yield an $X$-ray spectrum with a slope of $\beta_X =
-1.17^{-0.04}_{-0.03}$ and an $X$-ray flux decay of $\alpha_X = -0.9
\pm 0.3$ at 0.25 days after the burst.  Using the relationships of
\citet{sph1999} and \citet{cl1999} we can rule out the case where the
cooling frequency, $\nu_c$, is above the $X$-ray band since the
observed $\alpha_X$ and $\beta_X$ predict different values for the
electron index, $p$.  Therefore, the cooling break must be between the
lower edge of the \emph{Rossi-XTE\/} $X$-ray band (0.2 keV) and the
$R$ band at this time.  The optical and $X$-ray decay indices and
the $X$-ray spectral index at 0.25 days are consistent with $p = 2.2
\pm 0.1$, which is consistent with the observed decay index after the
jet break.

The spectral index computed in Section~4.1 for 0.65 days after the
burst predicts $p = 1.4$ if the cooling break frequency is below the
optical and $p = 2.4$ if it is above the optical.  Values for the
electron index of less than two represent infinite energy in the
electrons.  This strongly suggests that $\nu_c > \nu_{R}$ at this
time.  However, at $1.65 \le \Delta T \le 5.64$ days the optical
spectral slopes (see Section~4.1) are consistent with $\nu_c <
\nu_{R}$ and $p \approx 2$.  The case where $\nu_c > \nu_{R}$ at
these times implies $p \approx 3$, which is inconsistent with the
value of the electron index that was derived for $\Delta T = 0.25$
days.  \citet{tmg2003} used \emph{XMM-Newton\/} $X$-ray observations to
find $\beta_X = -0.92^{+0.26}_{-0.15}$ at 37 days and $\beta =
-1.1^{+0.4}_{-0.2}$ at 61 days.  Both of these are consistent with $p
\approx 2$ and $\nu_c < \nu_{R}$.  Therefore we believe that the
cooling break passed through the optical, moving toward radio
frequencies, between 0.65 and 1.65 days after the burst.  A cooling
frequency that decreases with time is the hallmark of a homogeneous
interstellar medium.  However, there may be local inhomogeneities on
scales that are small compared to the size of the fireball.

At $X$-ray frequencies interstellar absorption does not significantly
affect the slope of the spectrum, so the observed slope is a good
approximation of the actual slope.  Combining all of the $\beta_X$
values from \citet{tmg2003} yields $p = 2.2 \pm 0.1$. This predicts
that the optical spectrum has $\beta = -1.10 \pm 0.05$ at $\Delta T =
5.64$ days, which is close to the observed spectrum.  This
agreement strengthens our conclusion in Section~6.2 that there is no
strong evidence for dust in the host galaxy along the line of sight to
the burst.

\subsection{Separating the GRB from the Supernova}

To explore the nature of the supernova underlying the OT, we modeled
the spectrum as the sum of a power-law continuum and a peculiar Type
Ic SN.  Specifically, we chose for comparison SN~1998bw
\citep{patat01}, SN~1997ef \citep{iwamoto00}, and SN~2002ap (using our
own as yet unpublished spectra, but see, e.g., Kinugasa et al. 2002;
Foley et al. 2003).  We had 62 spectra of these three SNe, spanning
the epochs of seven days before maximum to several weeks past.  For
the power-law continuum, we chose to use one of our early spectra to
represent the afterglow of the GRB.  The spectrum at time $\Delta
T=5.80$ days was of high signal-to-noise ratio (S/N), and suffers from
little fringing at the red end.  Therefore, we smoothed this spectrum
to provide the fiducial power-law continuum of the OT for our model.
Other choices for the continuum did not affect our results
significantly.

\begin{figure}
\epsscale{0.75}
\rotatebox{90}{
\plotone{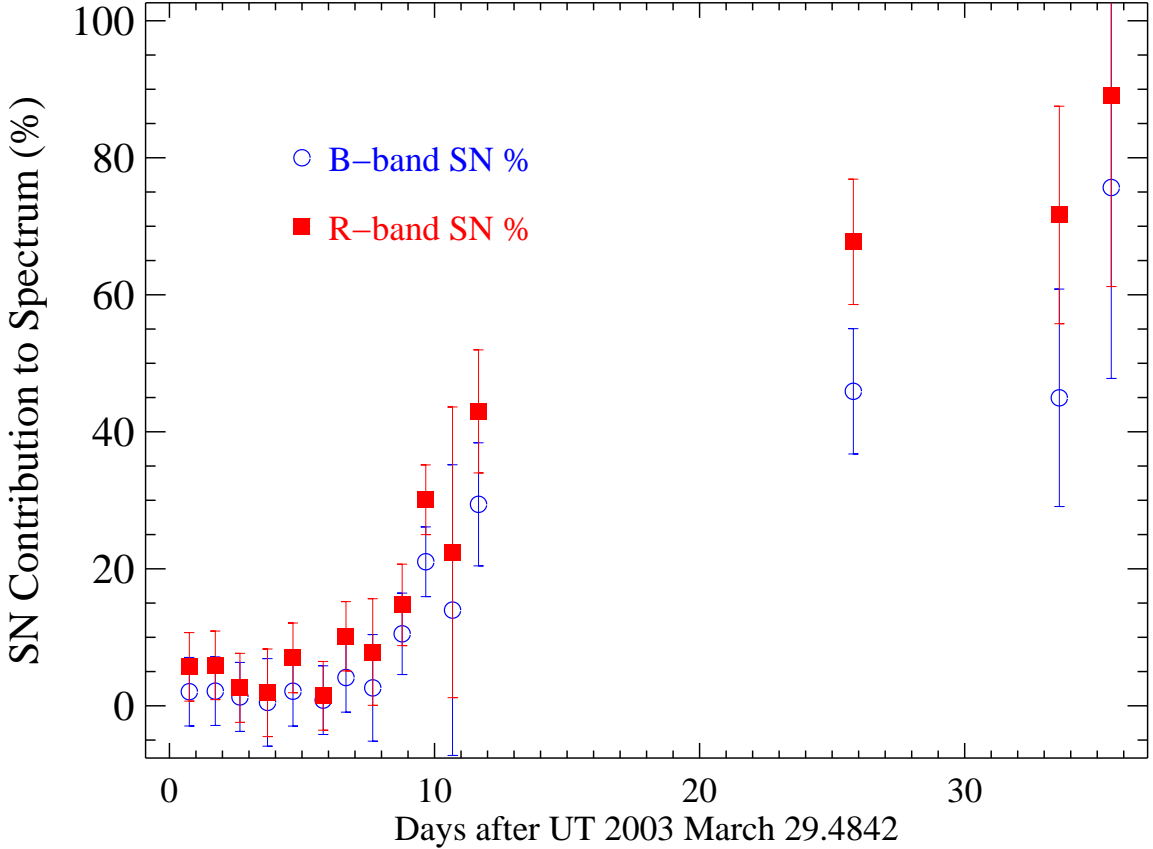}}
\caption{Relative contribution of a supernova spectrum to the
GRB~030329/SN~2003dh afterglow as a function of time in the $B$
(\emph{open circles}) and $R$ (\emph{filled squares}) bands.  Using
the technique described in the text, we derive a best fit to the
afterglow spectrum at each epoch with the fiducial power-law continuum
and the closest match from our set of peculiar SNe Ic.  We then
synthesize the relative $B$-band and $R$-band contributions.  There is
some scatter for the early epochs due to noise in the spectra, but a
clear deviation is evident starting at $\Delta T=7.67$ days, with a
subsequent rapid increase in the fraction of the overall spectrum
contributed by the SN.  Errors are estimated from the scatter when the
SN component is close to zero ($\Delta T < 6$ days) and from the scale
of the error in the least-squares minimization.
\label{snfrac}}
\end{figure}

To find the best match with a supernova spectrum, we compared each
spectrum of the afterglow with the sum of the fiducial continuum and a
spectrum of one of the SNe in the sample, both scaled in flux to match
the OT spectra.  We performed a least-squares fit, allowing the
fraction of continuum and SN to vary, finding the best combination of
continuum and SN for each of the SN spectra.  The minimum
least-squares deviation within this set was then taken as the best SN
match for that epoch of OT observation.  The results of the fits for
the spectra we modeled are listed in Table \ref{fittable}.  Figure
\ref{snfrac} shows the relative contribution to the OT spectrum by the
underlying SN in the $B$ and $R$ bands as a function of $\Delta T$.

Within the uncertainties of our fit, the SN fraction is consistent
with zero for the first few days after the GRB.  At $\Delta T=7.67$
days, the SN begins to appear in the spectrum, without strong evidence
for a supernova component before this.  Hjorth et al. (2003) report
evidence for the SN spectrum in their April 3 UT data ($\Delta T
\approx 4$ days), but we do not see any sign of a SN component at this
time.  There is a color change near $\Delta T \approx 5$ days as noted
above (see also Figure \ref{fig_spec}), but we attribute it to the
afterglow.  Our decomposition of the photometry into SN and afterglow
components (see below) indicates that, at most, the SN would have
contributed only a few percent of the total light at this point,
making it difficult to identify indisputable features.

Note that when the fit
indicates the presence of a supernova, the best match is almost always
SN~1998bw.  The only exceptions to this are from nights when the
spectrum of the OT are extremely noisy, implying that less weight
should be given to those results.  The least-squares deviation for the
spectra that do not match SN~1998bw is also much larger (see Table
\ref{fittable}).

\begin{figure}
\epsscale{0.8}
\rotatebox{90}{
\plotone{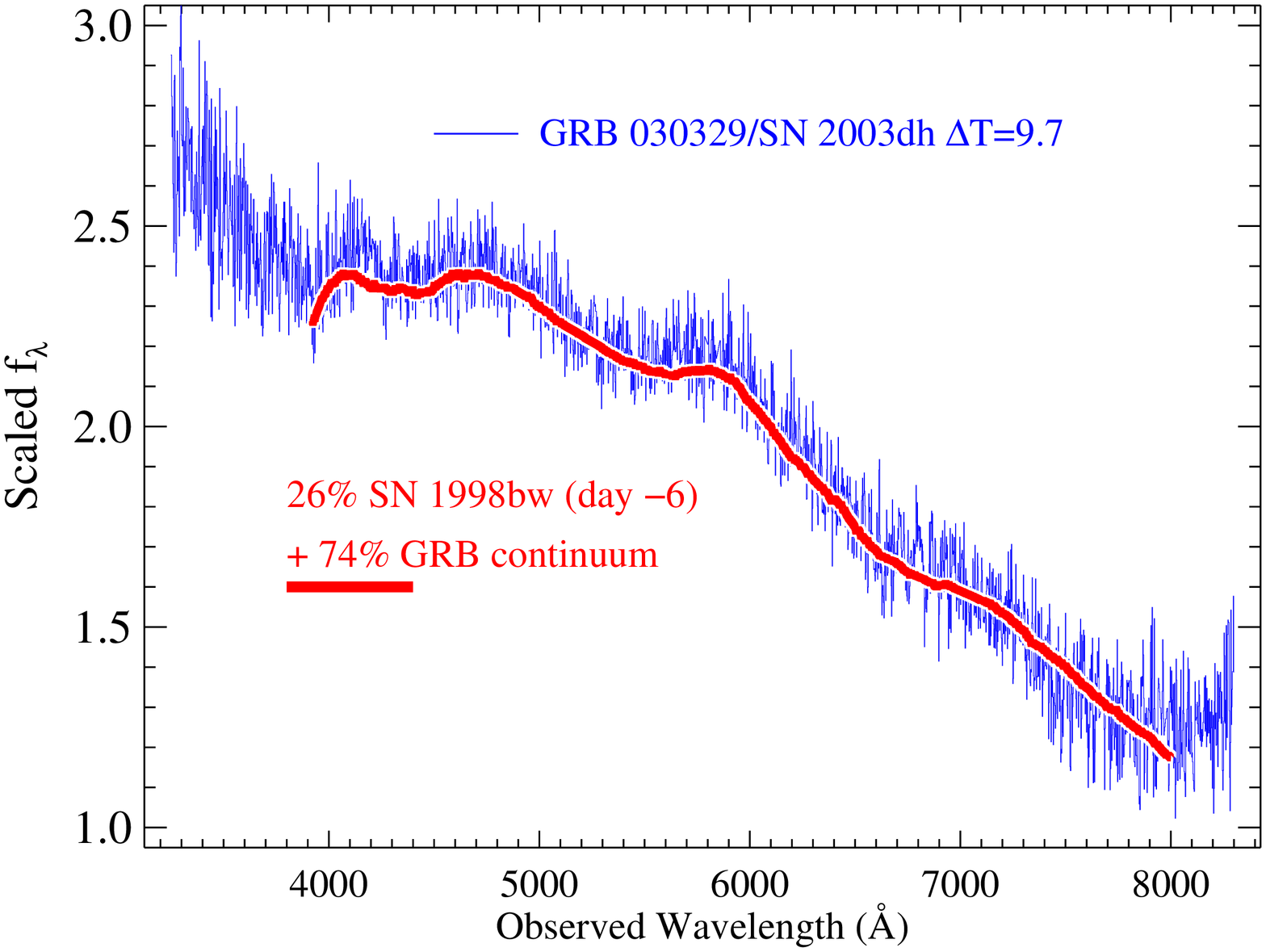}}
\caption{Observed spectrum (\emph{thin line}) of the
GRB~030329/SN~2003dh afterglow at $\Delta T=9.67$ days.  The model spectrum
(\emph{thick line}) consists of 74\% continuum and 26\% SN~1998bw from
6 days before maximum.  No other peculiar SN Ic spectrum provided as
good a fit.
\label{day09}}
\end{figure}

Our best spectrum (i.e., with the highest S/N) from this time when the
SN features begin to appear is at $\Delta T=9.67$ days.  In Figure
\ref{day09}, our best fit of 74\% continuum and 26\% SN~1998bw (at day
$-6$ relative to SN $B$-band maximum) is plotted over the observed
spectrum from this epoch.  The next-best fit is SN~1998bw at day $-$7.
Using a different early epoch to define the reference continuum does
not alter these results significantly.  It causes slight changes in
the relative percentages, but the same SN spectrum still produces the
best fit, albeit with a larger least-squares deviation.

\begin{figure}
\epsscale{0.8}
\rotatebox{90}{
\plotone{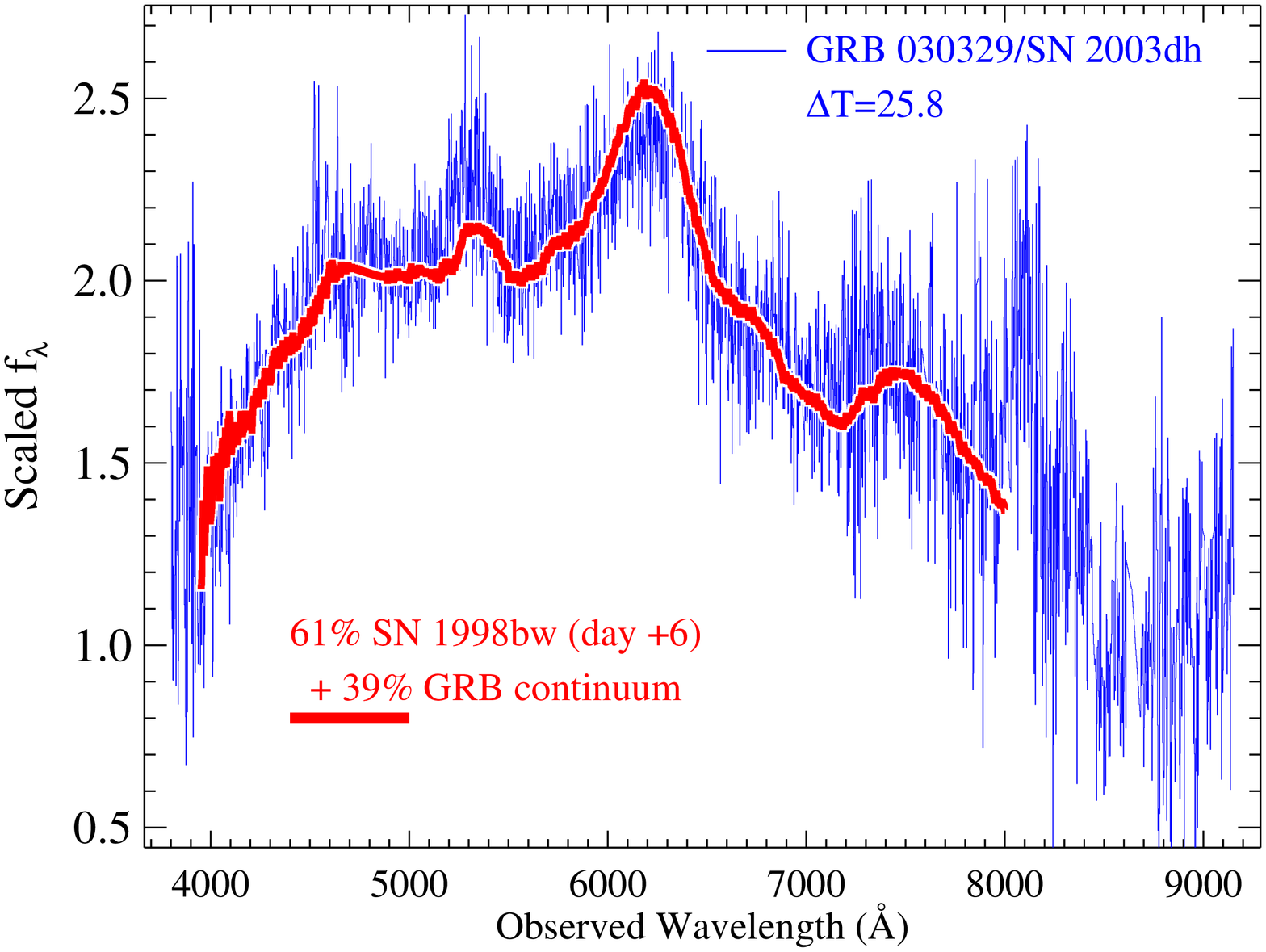}}
\caption{Observed spectrum (\emph{thin line}) of the
GRB~030329/SN~2003dh afterglow at $\Delta T=25.8$ days.  The model spectrum
(\emph{thick line}) consists of 39\% continuum and 61\% SN~1998bw from
6 days after maximum.
\label{day25}}
\end{figure}

\begin{figure}
\epsscale{0.8}
\rotatebox{90}{
\plotone{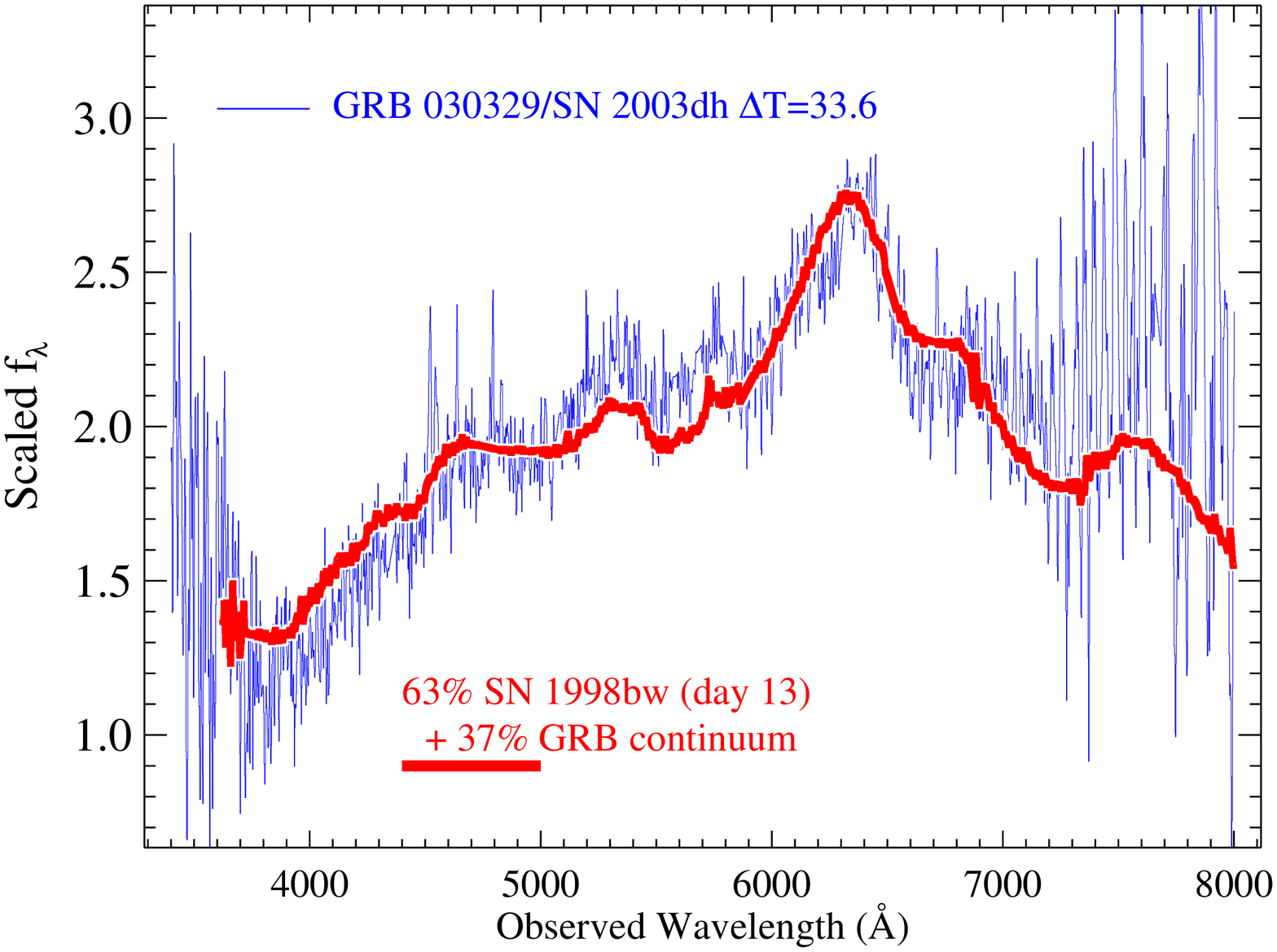}}
\caption{Observed spectrum (\emph{thin line}) of the
GRB~030329/SN~2003dh afterglow at $\Delta T=33.6$ days.  The model spectrum
(\emph{thick line}) consists of 37\% continuum and 63\% SN~1998bw from
13 days after maximum.
\label{day33}}
\end{figure}

The SN fraction contributing to the total spectrum increases steadily
with time.  By $\Delta T=25.8$ days, the SN fraction is $\sim$ 61\%, with
the best-fit SN being SN~1998bw at day +6 (Figure \ref{day25}).  The SN
percentage at $\Delta T=33.6$ days is still about 63\%, but the best match
is now SN~1998bw at day +13 (Figure \ref{day33}).  The rest-frame time
difference between $\Delta T=9.67$ days and $\Delta T=25.8$ days is 13.8 days
($z = 0.1685$).  For the best-fit SN spectra from those epochs,
SN~1998bw at day $-$6 and SN~1998bw at day +6 respectively, the time
difference is 12 days.  The rest-frame time difference between $\Delta
T=25.8$ days and $\Delta T=33.6$ days is 6.7 days, with a time difference
between the best-fit spectra for those epochs of 7 days.  The spectral
evolution determined from these fits indicates that SN~2003dh follows
SN~1998bw closely, and it is not as similar to SN~1997ef or
SN~2002ap.  The analysis by Kawabata et al. (2003) of their May 10
spectrum gives a phase for the spectrum of SN~2003dh that is
consistent with our dates, although they do consider SN~1997ef as a
viable alternative to SN~1998bw as a match for the SN component in the
afterglow.  

Once the spectrum of the SN has been separated from the power-law
continuum of the afterglow, one can consider the nature of SN~2003dh
itself.  The spectrum does not show any sign of broad hydrogen lines,
eliminating the Type II classification, nor is there the deep
\ion{Si}{2} $\lambda$6355 (usually blueshifted to $\sim$6150 \AA)
feature that is the hallmark of Type Ia SNe.  The optical helium line
absorptions that indicate SNe of Type Ib are not apparent either.
This leads to a classification of SN~2003dh as a Type Ic (see
Filippenko 1997 for a review of supernova classification).  Given the
striking correspondence with the Type Ic SN~1998bw shown above, this
is a natural classification for SN~2003dh.

The spectra of SN~1998bw (and other highly energetic SNe) are not
simple to interpret.  The high expansion velocities result in many
overlapping lines so that identification of specific line features is
problematic for the early phases of spectral evolution (see, e.g.,
Iwamoto et al. 1998; Stathakis et al. 2000; Nakamura et al. 2001;
Patat et al. 2001).  This includes spectra up to two weeks after
maximum, approximately the same epochs covered by our spectra of
SN~2003dh.  In fact, as Iwamoto et al. (1998) showed, the spectra at
these phases do not show line features.  The peaks in the spectra are
due to gaps in opacity, not individual spectral lines.  Detailed
modeling of the spectra can reveal some aspects of the composition of
the ejecta (e.g., Nakamura et al. 2001).  Such a model is beyond the
scope of this paper, but the spectra discussed herein and the spectrum
of Kawabata et al. (2003) are being analyzed for a future paper
(Mazzali et al., in preparation).

If the $\Delta T=9.67$ days spectrum for the afterglow does match
SN~1998bw at day $-$6, then limits can be placed on the timing of the
supernova explosion relative to the GRB.  The rest-frame time for
$\Delta T=9.67$ days is 8.2 days, implying that the time of the GRB
would correspond to $\sim$14 days before maximum for the SN.  The rise
times of SNe Ic are not well determined, especially for the small subset
of peculiar ones.  Stritzinger et al. (2002) found the rise time of
the Type Ib/c SN~1999ex was $\sim$18 days (in the $B$ band), while
Richmond et al. (1996) reported a rise time of $\sim$12 days (in the
$V$ band) for the Type Ic SN~1994I.  A rise time of $\sim$14 days for
SN~2003dh is certainly a reasonable number.  It also makes it
extremely unlikely that the SN exploded significantly earlier or later
than the time of the GRB, most likely within $\pm 2$ days of the GRB
itself.

The totality of data contained in this paper allows us to attempt to
decompose the light curve of the OT into the supernova and the
afterglow (power-law) component. From the spectral decomposition
procedure described above, we have the fraction of light in the
$BR$-bands for both components at various times, assuming that the
spectrum of the afterglow did not evolve since $\Delta T=5.64$ days.
As we find that the spectral evolution is remarkably close to that of
SN~1998bw, we model the $R$-band supernova component with the $V$-band
light curve of SN~1998bw (Galama et al. 1998a, b) stretched by
$(1+z)=1.1685$ and shifted in magnitude to obtain a good fit. The
afterglow component is fit by using the early points starting at
$\Delta T=5.64$ days with late points obtained via the spectral
decomposition. This can be done in both in the $B$ and in the $R$-band
and leads to consistent results, indicating that our assumption of the
afterglow not evolving in color at later times is indeed valid.

\begin{figure}
\plotone{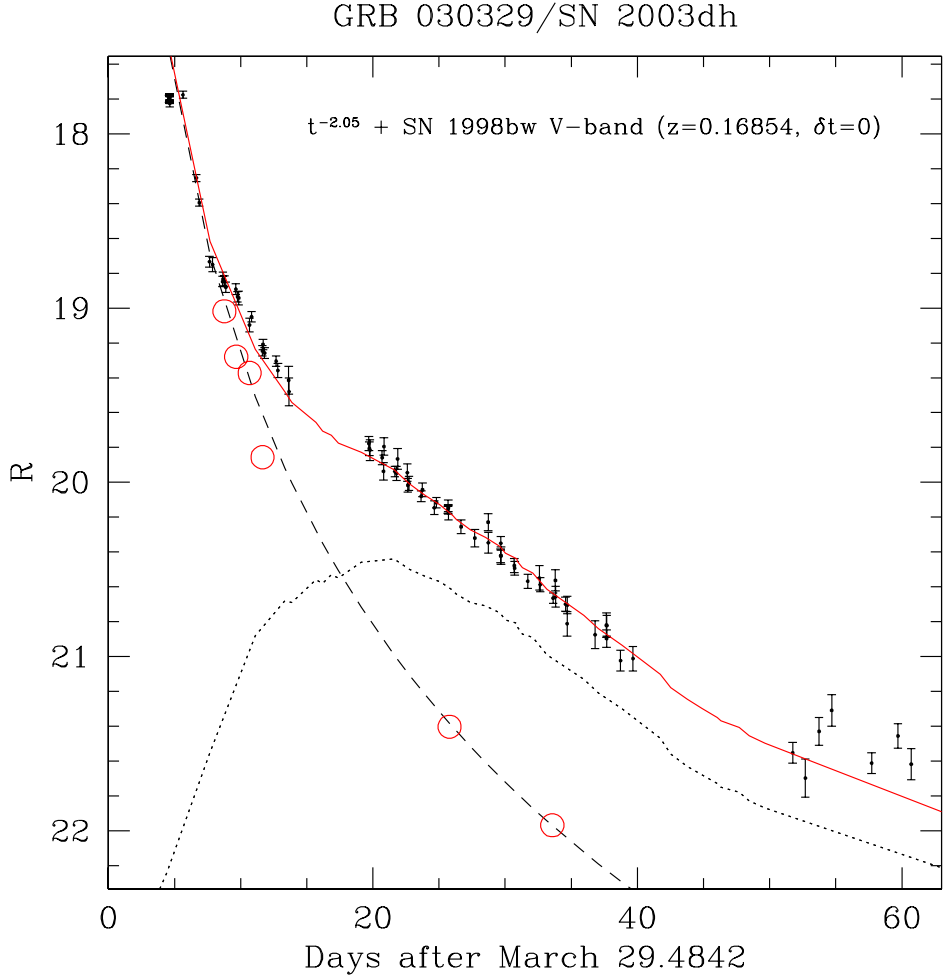}
\caption{Decomposition of the OT $R$-band light curve into the
supernova (\emph{dotted line}) and the power-law continuum
(\emph{dashed line}).  As the light curve model for the supernova, we
took the $V$-band light curve of SN~1998bw (Galama et al. 1998a, b)
stretched by $(1+z)=1.1685$ and shifted in magnitude.  The resulting
supernova light curve peaks at an apparent magnitude of $m_R=20.4$. No
offset in time has been applied between the GRB and the supernova. To
constrain the continuum, information from the spectral decomposition
was used (\emph{big open circles}).  See Section 6.4 for discussion.
\label{decomp}}
\end{figure}

The result of the decomposition of the OT $R$-band light curve into
the supernova and the power-law continuum is shown in Figure
\ref{decomp}.  The overall fit is remarkably good, given the
assumptions (such as using the stretched $V$-band light curve of
SN~1998bw as a proxy for the~SN 2003dh $R$-band light curve).  No time
offset between the supernova and the GRB was applied, and given how
good the fit is, we decided not to explore time offset as an
additional parameter. Introducing such an additional parameter would
most likely result in a somewhat better fit (indeed, we find that to
be the case for $\delta t\approx -2$ days), but this could easily be
an artifact with no physical significance, purely due to small
differences between SN~1998bw and SN~2003dh. At this point the
assumption that the GRB and the SN happened at the same time seems
most natural.

\subsection{The ``Jitter Episode''}

We also want to discuss briefly the ``Jitter Episode'' mentioned
earlier (Figure \ref{later}). Variations of $>30$\% on timescales of
$\sim 2 \;$days more than 50 days after the burst ($>40$ days in the
rest frame) are unlikely to be in the supernova component, as
such variations have never been observed in any other supernova. It is
much more likely that the afterglow of the GRB has exhibited another
episode of re-brightening, possibly due to interaction with
SN1987A-like rings ejected long ago from the progenitor.
Alternatively, the early afterglow had a complicated light curve,
possibly due to refreshed shocks (Granot, Nakar, \& Piran 2003), and
this late ``Jitter Episode'' could be somehow related to that earlier
behavior.  An extrapolation of the afterglow light implies that it was
more than a magnitude fainter than the supernova two months after the
burst so that the afterglow must have varied by nearly a factor of two
in brightness. A full investigation of that phenomenon is outside the
scope of the current paper, but we should note that its presence
complicates the search in the late light curve for the radioactive
decay component of the supernova, which could end up being masked by
the afterglow component. We continue observing this fascinating event
and will report future results in a paper by Bersier et al. (in
preparation).

\section{Summary}

We have presented optical and infrared photometry and optical
spectroscopy of \\
GRB~030329/SN~2003dh covering the first two months of
its evolution.  The early photometry shows a fairly complicated light
curve that cannot be simply fit in the manner of typical optical
afterglows.  Color changes are apparent in the early stages of the
afterglow, even before the supernova component begins to make a
contribution.  These color changes have been seen in other afterglows
(Matheson et al.~2003c; Bersier et al.~2003a), but the physical
mechanism that produces them is still a mystery.

At late times, the photometry becomes dominated by the SN component,
following the light curve of SN~1998bw fairly well.  The colors change
distinctly when the SN emerges, although from the light curves alone,
there is no clear bump from the SN as has been seen in higher-redshift
bursts.  There is the ``Jitter Episode'' near the two-month mark,
indicating that the afterglow may still contribute significantly to
the observed brightness even at this late date.
 
The evidence from photometry alone would not be a completely
convincing case for the presence of a supernova.  The spectra,
especially the day-by-day coverage for the first twelve days of the
burst, show the transition from the power-law continuum
of the afterglow to the broad features characteristic of a supernova.
By subtracting off the continuum, the SN becomes directly apparent,
and the correspondence with SN~1998bw at virtually all of the epochs
for which we have spectroscopy is striking.  Taking into account the
cosmological time dilation, the development of SN~2003dh follows that
of SN~1998bw almost exactly.  Using the spectroscopic decomposition of
our data, we can separate the light curve into afterglow and SN
components, again showing that SN~2003dh follows SN~1998bw. The
decomposition suggests that the supernova explosion occurred close
to the time of the gamma-ray burst.

The spectroscopy of the optical afterglow of GRB~030329, as first
shown by Stanek et al. (2003a), provided direct evidence that at least
some of the long-burst GRBs are related to core-collapse SNe.  We have
shown with a larger set of data that the SN component is similar
to SN~1998bw, an unusual Type Ic SN.  It is not clear yet whether all
long-burst GRBs arise from SNe.  Catching another GRB at a redshift
this low is unlikely, but large telescopes may be able to discern SNe
in some of the relatively nearby bursts.  With this one example,
though, we now have solid evidence that some GRBs and SNe have the
same progenitors.

\acknowledgments

We would like to thank the staffs of the MMT, FLWO, Las Campanas,
Lick, Keck, and Kitt Peak National Observatories.  We are grateful to
J.~McAfee and A.~Milone for helping with the MMT spectra.  We thank
the \emph{HETE2} team, Scott Barthelmy, and the GRB Coordinates Network
(GCN) for the quick turnaround in providing precise GRB positions to
the astronomical community.  We also thank all the observers who
provided their data and analysis through the GCN.  KPNO, a part of the
National Optical Astronomy Observatory, is operated by the Association
of Universities for Research in Astronomy, Inc. (AURA), under a
cooperative agreement with the National Science Foundation.  PMG and
STH acknowledge the support of NASA/LTSA grant NAG5-9364.  DJE was
supported by NSF grant AST-0098577 and by an Alfred P. Sloan Research
Fellowship. DM, LI and JM are supported by FONDAP Center for
Astrophysics 15010003. JK was supported by KBN grant 5P03D004.21. JCL
acknowledges financial support from NSF grant AST-9900789. PM was
supported by a Carnegie Starr Fellowship. EWO was partially supported
by NSF grant AST 0098518. BP is supported by NASA grant NAG5-9274. RAW
and RAJ are supported by NASA Grant NAG5-12460.  DCL acknowledges
financial support provided by NASA through grant GO-9155 from the
Space Telescope Science Institute, which is operated by the
Association of Universities for Research in Astronomy, Inc., under
NASA contract NAS 5-26555.  The work of AVF's group at the University
of California, Berkeley, is supported by NSF grant AST-0307894, as
well as by the Sylvia and Jim Katzman Foundation.  KAIT was made
possible by generous donations from Sun Microsystems, Inc., the
Hewlett-Packard Company, AutoScope Corporation, Lick Observatory, the
NSF, the University of California, and the Katzman Foundation.

\begin{deluxetable}{ r c c c l }
\tabletypesize{\small}
\tablewidth{0pt}
\tablecaption{JOURNAL OF PHOTOMETRIC OBSERVATIONS\label{phottable}}
\tablehead{\colhead{$\Delta$T\tablenotemark{a}} &
\colhead{Mag} &
\colhead{$\sigma_m$} &
\colhead{Filter} &
\colhead{Observatory\tablenotemark{b}}
}
\startdata
 0.6494 & 15.296 & 0.020 & U & FLWO \\ 
 0.6601 & 15.346 & 0.020 & U & FLWO \\ 
 0.6796 & 15.394 & 0.050 & U & KAIT \\ 
 0.6841 & 15.381 & 0.020 & U & FLWO \\ 
 0.6989 & 15.369 & 0.010 & U & FLWO \\ 
 0.7041 & 15.442 & 0.040 & U & KAIT \\ 
 0.7091 & 15.434 & 0.010 & U & FLWO \\ 
 0.7191 & 15.490 & 0.010 & U & FLWO \\ 
 0.7198 & 15.458 & 0.050 & U & KAIT \\ 
 0.7291 & 15.525 & 0.010 & U & FLWO \\ 
 0.7391 & 15.557 & 0.010 & U & FLWO \\ 
 0.7505 & 15.584 & 0.010 & U & FLWO \\ 
 0.7590 & 15.599 & 0.050 & U & KAIT \\ 
 0.7609 & 15.610 & 0.010 & U & FLWO \\ 
 0.7712 & 15.630 & 0.010 & U & FLWO \\ 
 0.7746 & 15.624 & 0.040 & U & KAIT \\ 
 0.7816 & 15.643 & 0.010 & U & FLWO \\ 
 0.7920 & 15.678 & 0.010 & U & FLWO \\ 
 0.8023 & 15.715 & 0.010 & U & FLWO \\ 
 0.8098 & 15.707 & 0.050 & U & KAIT \\ 
 0.8127 & 15.758 & 0.020 & U & FLWO \\ 
 0.8231 & 15.774 & 0.020 & U & FLWO \\ 
 0.8324 & 15.814 & 0.050 & U & KAIT \\ 
 0.8334 & 15.823 & 0.020 & U & FLWO \\ 
 0.8438 & 15.775 & 0.020 & U & FLWO \\ 
 0.8536 & 15.851 & 0.060 & U & KAIT \\ 
 0.8542 & 15.850 & 0.020 & U & FLWO \\ 
 0.8645 & 15.862 & 0.020 & U & FLWO \\ 
 0.8694 & 15.993 & 0.060 & U & KAIT \\ 
 0.8749 & 15.941 & 0.020 & U & FLWO \\ 
 0.8853 & 15.936 & 0.060 & U & KAIT \\ 
 0.8853 & 15.963 & 0.020 & U & FLWO \\ 
 0.8956 & 15.982 & 0.020 & U & FLWO \\ 
 0.9039 & 16.025 & 0.060 & U & KAIT \\ 
 0.9059 & 16.059 & 0.020 & U & FLWO \\ 
 0.9164 & 16.063 & 0.020 & U & FLWO \\ 
 0.9199 & 16.086 & 0.070 & U & KAIT \\ 
 0.9268 & 16.072 & 0.020 & U & FLWO \\ 
 0.9360 & 16.087 & 0.080 & U & KAIT \\ 
 0.9371 & 16.118 & 0.020 & U & FLWO \\ 
 0.9475 & 16.155 & 0.030 & U & FLWO \\ 
 0.9578 & 16.176 & 0.030 & U & FLWO \\ 
 1.6398 & 16.415 & 0.020 & U & FLWO \\ 
 1.7351 & 16.539 & 0.020 & U & FLWO \\ 
 1.7852 & 16.546 & 0.060 & U & KAIT \\ 
 1.8337 & 16.681 & 0.020 & U & FLWO \\ 
 1.9092 & 16.771 & 0.020 & U & FLWO \\ 
 2.8438 & 17.282 & 0.020 & U & FLWO \\ 
 2.9549 & 17.329 & 0.030 & U & FLWO \\ 
 3.7822 & 17.578 & 0.030 & U & FLWO \\ 
 3.8012 & 17.610 & 0.030 & U & FLWO \\ 
 3.8262 & 17.670 & 0.030 & U & FLWO \\ 
 3.8454 & 17.704 & 0.030 & U & FLWO \\ 
 4.6574 & 18.060 & 0.030 & U & FLWO \\ 
 5.6493 & 18.011 & 0.040 & U & FLWO \\ 
 6.6648 & 18.664 & 0.060 & U & FLWO \\ 
 0.6411 & 15.903 & 0.010 & B & FLWO \\ 
 0.6518 & 15.921 & 0.010 & B & FLWO \\ 
 0.6663 & 15.977 & 0.010 & B & FLWO \\ 
 0.6836 & 16.013 & 0.020 & B & KAIT \\ 
 0.6863 & 16.033 & 0.010 & B & FLWO \\ 
 0.6912 & 16.036 & 0.010 & B & FLWO \\ 
 0.7019 & 16.085 & 0.010 & B & FLWO \\ 
 0.7080 & 16.076 & 0.020 & B & KAIT \\ 
 0.7120 & 16.100 & 0.010 & B & FLWO \\ 
 0.7220 & 16.146 & 0.010 & B & FLWO \\ 
 0.7238 & 16.117 & 0.010 & B & KAIT \\ 
 0.7320 & 16.185 & 0.010 & B & FLWO \\ 
 0.7432 & 16.161 & 0.010 & B & FLWO \\ 
 0.7536 & 16.212 & 0.010 & B & FLWO \\ 
 0.7629 & 16.229 & 0.010 & B & KAIT \\ 
 0.7639 & 16.242 & 0.010 & B & FLWO \\ 
 0.7743 & 16.286 & 0.010 & B & FLWO \\ 
 0.7786 & 16.288 & 0.020 & B & KAIT \\ 
 0.7847 & 16.302 & 0.010 & B & FLWO \\ 
 0.7950 & 16.328 & 0.010 & B & FLWO \\ 
 0.8054 & 16.346 & 0.010 & B & FLWO \\ 
 0.8138 & 16.374 & 0.030 & B & KAIT \\ 
 0.8157 & 16.379 & 0.010 & B & FLWO \\ 
 0.8262 & 16.386 & 0.010 & B & FLWO \\ 
 0.8363 & 16.400 & 0.030 & B & KAIT \\ 
 0.8365 & 16.429 & 0.010 & B & FLWO \\ 
 0.8468 & 16.444 & 0.010 & B & FLWO \\ 
 0.8572 & 16.491 & 0.010 & B & FLWO \\ 
 0.8576 & 16.505 & 0.040 & B & KAIT \\ 
 0.8676 & 16.519 & 0.010 & B & FLWO \\ 
 0.8734 & 16.505 & 0.030 & B & KAIT \\ 
 0.8780 & 16.542 & 0.010 & B & FLWO \\ 
 0.8883 & 16.562 & 0.020 & B & FLWO \\ 
 0.8893 & 16.531 & 0.050 & B & KAIT \\ 
 0.8986 & 16.626 & 0.020 & B & FLWO \\ 
 0.9079 & 16.595 & 0.070 & B & KAIT \\ 
 0.9090 & 16.640 & 0.020 & B & FLWO \\ 
 0.9195 & 16.654 & 0.020 & B & FLWO \\ 
 0.9239 & 16.676 & 0.030 & B & KAIT \\ 
 0.9298 & 16.665 & 0.020 & B & FLWO \\ 
 0.9399 & 16.710 & 0.020 & B & KAIT \\ 
 0.9402 & 16.724 & 0.020 & B & FLWO \\ 
 0.9505 & 16.748 & 0.020 & B & FLWO \\ 
 1.6447 & 17.002 & 0.010 & B & FLWO \\ 
 1.7399 & 17.132 & 0.010 & B & FLWO \\ 
 1.7909 & 17.210 & 0.020 & B & KAIT \\ 
 1.8385 & 17.267 & 0.010 & B & FLWO \\ 
 1.9144 & 17.331 & 0.010 & B & FLWO \\ 
 2.8485 & 17.805 & 0.020 & B & FLWO \\ 
 2.9596 & 17.937 & 0.020 & B & FLWO \\ 
 3.7261 & 18.104 & 0.040 & B & FLWO \\ 
 3.7875 & 18.154 & 0.020 & B & FLWO \\ 
 3.8125 & 18.197 & 0.020 & B & FLWO \\ 
 3.8316 & 18.221 & 0.020 & B & FLWO \\ 
 3.8621 & 18.244 & 0.020 & B & FLWO \\ 
 4.6437 & 18.632 & 0.020 & B & FLWO \\ 
 5.6356 & 18.528 & 0.030 & B & FLWO \\ 
 5.7063 & 18.548 & 0.040 & B & LCO100 \\ 
 5.7100 & 18.569 & 0.040 & B & LCO100 \\ 
 6.5796 & 19.076 & 0.040 & B & LCO100 \\ 
 6.5840 & 19.069 & 0.040 & B & LCO100 \\ 
 6.6511 & 19.053 & 0.030 & B & FLWO \\ 
 6.8690 & 19.149 & 0.030 & B & FLWO \\ 
 7.6383 & 19.478 & 0.060 & B & FLWO \\ 
 7.8741 & 19.632 & 0.050 & B & FLWO \\ 
 8.6528 & 19.699 & 0.060 & B & FLWO \\ 
 8.7897 & 19.774 & 0.040 & B & FLWO \\ 
 8.8919 & 19.698 & 0.040 & B & FLWO \\ 
 9.6497 & 19.750 & 0.050 & B & FLWO \\ 
 9.7762 & 19.871 & 0.040 & B & FLWO \\ 
 9.8720 & 19.820 & 0.060 & B & FLWO \\ 
10.7558 & 19.953 & 0.040 & B & FLWO \\ 
11.7358 & 20.112 & 0.040 & B & FLWO \\ 
12.7403 & 20.394 & 0.050 & B & FLWO \\ 
21.7000 & 21.296 & 0.060 & B & FLWO \\ 
22.6667 & 21.241 & 0.070 & B & FLWO \\ 
23.6814 & 21.382 & 0.070 & B & FLWO \\ 
24.6858 & 21.440 & 0.060 & B & FLWO \\ 
25.6888 & 21.671 & 0.040 & B & KPNO4m \\ 
25.6968 & 21.583 & 0.040 & B & KPNO4m \\ 
25.7402 & 21.516 & 0.060 & B & FLWO \\ 
26.6888 & 21.675 & 0.040 & B & KPNO4m \\ 
26.6968 & 21.660 & 0.040 & B & KPNO4m \\ 
26.7191 & 21.644 & 0.080 & B & FLWO \\ 
27.6718 & 21.714 & 0.030 & B & KPNO4m \\ 
27.6878 & 21.725 & 0.020 & B & KPNO4m \\ 
28.6688 & 21.794 & 0.030 & B & KPNO4m \\ 
29.6408 & 21.869 & 0.060 & B & KPNO4m \\ 
29.6528 & 21.900 & 0.020 & B & KPNO4m \\ 
30.6808 & 21.971 & 0.030 & B & KPNO4m \\ 
31.6678 & 22.034 & 0.020 & B & KPNO4m \\ 
31.6748 & 22.023 & 0.020 & B & KPNO4m \\ 
32.6818 & 22.066 & 0.040 & B & KPNO4m \\ 
33.6558 & 22.124 & 0.030 & B & KPNO4m \\ 
33.6628 & 22.173 & 0.050 & B & KPNO4m \\ 
34.6538 & 22.121 & 0.050 & B & KPNO4m \\ 
34.6608 & 22.185 & 0.040 & B & KPNO4m \\ 
36.5578 & 22.268 & 0.040 & B & LCO100 \\ 
37.5148 & 22.285 & 0.040 & B & LCO100 \\ 
37.6718 & 22.254 & 0.040 & B & KPNO4m \\ 
37.6808 & 22.357 & 0.040 & B & KPNO4m \\ 
38.5538 & 22.273 & 0.040 & B & LCO100 \\ 
39.5468 & 22.335 & 0.060 & B & LCO100 \\ 
 0.6432 & 15.545 & 0.010 & V & FLWO \\ 
 0.6539 & 15.574 & 0.010 & V & FLWO \\ 
 0.6709 & 15.614 & 0.010 & V & FLWO \\ 
 0.6876 & 15.729 & 0.030 & V & KAIT \\ 
 0.6927 & 15.692 & 0.010 & V & FLWO \\ 
 0.7036 & 15.720 & 0.010 & V & FLWO \\ 
 0.7111 & 15.770 & 0.010 & V & KAIT \\ 
 0.7136 & 15.750 & 0.010 & V & FLWO \\ 
 0.7236 & 15.784 & 0.010 & V & FLWO \\ 
 0.7269 & 15.807 & 0.010 & V & KAIT \\ 
 0.7336 & 15.809 & 0.010 & V & FLWO \\ 
 0.7450 & 15.822 & 0.010 & V & FLWO \\ 
 0.7554 & 15.851 & 0.010 & V & FLWO \\ 
 0.7657 & 15.893 & 0.010 & V & FLWO \\ 
 0.7660 & 15.931 & 0.030 & V & KAIT \\ 
 0.7761 & 15.911 & 0.010 & V & FLWO \\ 
 0.7817 & 15.965 & 0.030 & V & KAIT \\ 
 0.7865 & 15.946 & 0.010 & V & FLWO \\ 
 0.7968 & 15.962 & 0.010 & V & FLWO \\ 
 0.8072 & 15.977 & 0.010 & V & FLWO \\ 
 0.8176 & 16.034 & 0.010 & V & FLWO \\ 
 0.8178 & 16.070 & 0.020 & V & KAIT \\ 
 0.8280 & 16.051 & 0.010 & V & FLWO \\ 
 0.8383 & 16.069 & 0.010 & V & FLWO \\ 
 0.8395 & 16.139 & 0.020 & V & KAIT \\ 
 0.8486 & 16.104 & 0.010 & V & FLWO \\ 
 0.8590 & 16.155 & 0.010 & V & FLWO \\ 
 0.8607 & 16.192 & 0.020 & V & KAIT \\ 
 0.8694 & 16.151 & 0.010 & V & FLWO \\ 
 0.8765 & 16.239 & 0.030 & V & KAIT \\ 
 0.8798 & 16.194 & 0.010 & V & FLWO \\ 
 0.8901 & 16.208 & 0.020 & V & FLWO \\ 
 0.8924 & 16.254 & 0.050 & V & KAIT \\ 
 0.9004 & 16.247 & 0.020 & V & FLWO \\ 
 0.9109 & 16.245 & 0.020 & V & FLWO \\ 
 0.9110 & 16.269 & 0.070 & V & KAIT \\ 
 0.9213 & 16.309 & 0.020 & V & FLWO \\ 
 0.9270 & 16.360 & 0.050 & V & KAIT \\ 
 0.9316 & 16.331 & 0.020 & V & FLWO \\ 
 0.9420 & 16.336 & 0.020 & V & FLWO \\ 
 0.9432 & 16.395 & 0.040 & V & KAIT \\ 
 0.9523 & 16.366 & 0.020 & V & FLWO \\ 
 1.6482 & 16.628 & 0.010 & V & FLWO \\ 
 1.7435 & 16.737 & 0.010 & V & FLWO \\ 
 1.7949 & 16.861 & 0.030 & V & KAIT \\ 
 1.8421 & 16.859 & 0.010 & V & FLWO \\ 
 1.8564 & 16.906 & 0.070 & V & KAIT \\ 
 1.9179 & 16.925 & 0.010 & V & FLWO \\ 
 2.8520 & 17.416 & 0.010 & V & FLWO \\ 
 2.9632 & 17.486 & 0.020 & V & FLWO \\ 
 3.7296 & 17.717 & 0.050 & V & FLWO \\ 
 3.7720 & 17.748 & 0.020 & V & FLWO \\ 
 3.7911 & 17.757 & 0.020 & V & FLWO \\ 
 3.8160 & 17.783 & 0.020 & V & FLWO \\ 
 3.8352 & 17.825 & 0.020 & V & FLWO \\ 
 3.8657 & 17.814 & 0.020 & V & FLWO \\ 
 4.6472 & 18.110 & 0.020 & V & FLWO \\ 
 5.6391 & 18.110 & 0.020 & V & FLWO \\ 
 5.7150 & 18.099 & 0.020 & V & LCO100 \\ 
 5.7177 & 18.072 & 0.020 & V & LCO100 \\ 
 6.5882 & 18.585 & 0.020 & V & LCO100 \\ 
 6.5908 & 18.584 & 0.020 & V & LCO100 \\ 
 6.6546 & 18.617 & 0.020 & V & FLWO \\ 
 6.8737 & 18.716 & 0.020 & V & FLWO \\ 
 7.6431 & 19.064 & 0.040 & V & FLWO \\ 
 7.8788 & 19.024 & 0.040 & V & FLWO \\ 
 8.6401 & 19.158 & 0.060 & V & FLWO \\ 
 8.7770 & 19.225 & 0.040 & V & FLWO \\ 
 8.8792 & 19.227 & 0.040 & V & FLWO \\ 
 9.6370 & 19.170 & 0.060 & V & FLWO \\ 
 9.7635 & 19.206 & 0.030 & V & FLWO \\ 
 9.8592 & 19.194 & 0.050 & V & FLWO \\ 
10.6658 & 19.452 & 0.050 & V & FLWO \\ 
10.8158 & 19.458 & 0.040 & V & FLWO \\ 
11.6590 & 19.575 & 0.040 & V & FLWO \\ 
11.7723 & 19.570 & 0.040 & V & FLWO \\ 
25.6828 & 20.632 & 0.040 & V & KPNO4m \\ 
25.7038 & 20.620 & 0.030 & V & KPNO4m \\ 
26.6778 & 20.701 & 0.040 & V & KPNO4m \\ 
26.6838 & 20.680 & 0.040 & V & KPNO4m \\ 
27.6668 & 20.812 & 0.020 & V & KPNO4m \\ 
27.6828 & 20.815 & 0.020 & V & KPNO4m \\ 
28.6748 & 20.885 & 0.020 & V & KPNO4m \\ 
29.6358 & 20.953 & 0.040 & V & KPNO4m \\ 
29.6468 & 20.948 & 0.020 & V & KPNO4m \\ 
30.6878 & 21.058 & 0.020 & V & KPNO4m \\ 
31.6558 & 21.080 & 0.020 & V & KPNO4m \\ 
31.6618 & 21.090 & 0.020 & V & KPNO4m \\ 
32.6888 & 21.122 & 0.030 & V & KPNO4m \\ 
33.6428 & 21.210 & 0.030 & V & KPNO4m \\ 
33.6488 & 21.224 & 0.020 & V & KPNO4m \\ 
33.6818 & 21.160 & 0.040 & V & KPNO4m \\ 
34.6468 & 21.337 & 0.050 & V & KPNO4m \\ 
34.6698 & 21.310 & 0.050 & V & KPNO4m \\ 
36.5368 & 21.374 & 0.030 & V & LCO100 \\ 
36.6688 & 21.392 & 0.030 & V & KPNO4m \\ 
36.6768 & 21.360 & 0.030 & V & KPNO4m \\ 
37.4928 & 21.433 & 0.030 & V & LCO100 \\ 
38.5468 & 21.487 & 0.030 & V & LCO100 \\ 
39.5328 & 21.468 & 0.030 & V & LCO100 \\ 
 0.5678 & 15.021 & 0.020 & R & Clay \\ 
 0.5698 & 14.990 & 0.020 & R & Clay \\ 
 0.6251 & 15.182 & 0.010 & R & FLWO \\ 
 0.6276 & 15.165 & 0.010 & R & FLWO \\ 
 0.6307 & 15.180 & 0.010 & R & FLWO \\ 
 0.6452 & 15.228 & 0.010 & R & FLWO \\ 
 0.6559 & 15.266 & 0.010 & R & FLWO \\ 
 0.6755 & 15.306 & 0.010 & R & FLWO \\ 
 0.6920 & 15.400 & 0.030 & R & KAIT \\ 
 0.6943 & 15.381 & 0.010 & R & FLWO \\ 
 0.7049 & 15.387 & 0.010 & R & FLWO \\ 
 0.7133 & 15.424 & 0.020 & R & KAIT \\ 
 0.7150 & 15.422 & 0.010 & R & FLWO \\ 
 0.7250 & 15.464 & 0.010 & R & FLWO \\ 
 0.7291 & 15.463 & 0.010 & R & KAIT \\ 
 0.7350 & 15.502 & 0.010 & R & FLWO \\ 
 0.7464 & 15.504 & 0.010 & R & FLWO \\ 
 0.7567 & 15.535 & 0.010 & R & FLWO \\ 
 0.7671 & 15.576 & 0.010 & R & FLWO \\ 
 0.7683 & 15.594 & 0.020 & R & KAIT \\ 
 0.7775 & 15.578 & 0.010 & R & FLWO \\ 
 0.7839 & 15.635 & 0.020 & R & KAIT \\ 
 0.7878 & 15.624 & 0.010 & R & FLWO \\ 
 0.7982 & 15.634 & 0.010 & R & FLWO \\ 
 0.8086 & 15.664 & 0.010 & R & FLWO \\ 
 0.8190 & 15.702 & 0.010 & R & FLWO \\ 
 0.8218 & 15.700 & 0.010 & R & KAIT \\ 
 0.8293 & 15.750 & 0.010 & R & FLWO \\ 
 0.8396 & 15.742 & 0.010 & R & FLWO \\ 
 0.8417 & 15.785 & 0.080 & R & KAIT \\ 
 0.8500 & 15.774 & 0.010 & R & FLWO \\ 
 0.8604 & 15.802 & 0.010 & R & FLWO \\ 
 0.8630 & 15.801 & 0.010 & R & KAIT \\ 
 0.8708 & 15.852 & 0.010 & R & FLWO \\ 
 0.8787 & 15.875 & 0.010 & R & KAIT \\ 
 0.8811 & 15.875 & 0.010 & R & FLWO \\ 
 0.8915 & 15.885 & 0.010 & R & FLWO \\ 
 0.8947 & 15.883 & 0.020 & R & KAIT \\ 
 0.9018 & 15.909 & 0.010 & R & FLWO \\ 
 0.9123 & 15.946 & 0.010 & R & FLWO \\ 
 0.9132 & 15.924 & 0.020 & R & KAIT \\ 
 0.9226 & 15.983 & 0.020 & R & FLWO \\ 
 0.9293 & 15.941 & 0.020 & R & KAIT \\ 
 0.9330 & 16.002 & 0.020 & R & FLWO \\ 
 0.9433 & 16.034 & 0.020 & R & FLWO \\ 
 0.9454 & 16.078 & 0.060 & R & KAIT \\ 
 0.9537 & 16.069 & 0.020 & R & FLWO \\ 
 1.5838 & 16.233 & 0.020 & R & Clay \\ 
 1.5848 & 16.230 & 0.020 & R & Clay \\ 
 1.5858 & 16.225 & 0.020 & R & Clay \\ 
 1.5868 & 16.227 & 0.020 & R & Clay \\ 
 1.6514 & 16.271 & 0.010 & R & FLWO \\ 
 1.7466 & 16.403 & 0.010 & R & FLWO \\ 
 1.7989 & 16.536 & 0.020 & R & KAIT \\ 
 1.8452 & 16.541 & 0.010 & R & FLWO \\ 
 1.8611 & 16.647 & 0.070 & R & KAIT \\ 
 1.9211 & 16.575 & 0.010 & R & FLWO \\ 
 2.5778 & 16.815 & 0.020 & R & Clay \\ 
 2.5788 & 16.849 & 0.020 & R & Clay \\ 
 2.5808 & 16.858 & 0.020 & R & Clay \\ 
 2.5818 & 16.856 & 0.020 & R & Clay \\ 
 2.8552 & 17.069 & 0.010 & R & FLWO \\ 
 2.9663 & 17.152 & 0.010 & R & FLWO \\ 
 3.5868 & 17.241 & 0.020 & R & Clay \\ 
 3.5968 & 17.245 & 0.020 & R & Clay \\ 
 3.5978 & 17.254 & 0.020 & R & Clay \\ 
 3.5988 & 17.242 & 0.020 & R & Clay \\ 
 3.5998 & 17.252 & 0.020 & R & Clay \\ 
 3.6058 & 17.230 & 0.020 & R & Clay \\ 
 3.7328 & 17.361 & 0.030 & R & FLWO \\ 
 3.7752 & 17.387 & 0.020 & R & FLWO \\ 
 3.7942 & 17.408 & 0.010 & R & FLWO \\ 
 3.8192 & 17.433 & 0.010 & R & FLWO \\ 
 3.8383 & 17.462 & 0.010 & R & FLWO \\ 
 3.8688 & 17.491 & 0.020 & R & FLWO \\ 
 4.6008 & 17.799 & 0.020 & R & Clay \\ 
 4.6048 & 17.788 & 0.020 & R & Clay \\ 
 4.6058 & 17.788 & 0.020 & R & Clay \\ 
 4.6068 & 17.795 & 0.020 & R & Clay \\ 
 4.6088 & 17.797 & 0.020 & R & Clay \\ 
 4.6098 & 17.803 & 0.020 & R & Clay \\ 
 4.6348 & 17.807 & 0.020 & R & Clay \\ 
 4.6368 & 17.803 & 0.020 & R & Clay \\ 
 4.6388 & 17.826 & 0.020 & R & Clay \\ 
 4.6504 & 17.790 & 0.020 & R & FLWO \\ 
 5.6423 & 17.776 & 0.020 & R & FLWO \\ 
 6.6578 & 18.253 & 0.020 & R & FLWO \\ 
 6.8776 & 18.395 & 0.020 & R & FLWO \\ 
 7.6469 & 18.734 & 0.030 & R & FLWO \\ 
 7.8827 & 18.750 & 0.040 & R & FLWO \\ 
 8.6440 & 18.848 & 0.030 & R & FLWO \\ 
 8.6653 & 18.834 & 0.040 & R & FLWO \\ 
 8.7808 & 18.845 & 0.030 & R & FLWO \\ 
 8.8830 & 18.881 & 0.030 & R & FLWO \\ 
 9.6409 & 18.891 & 0.030 & R & FLWO \\ 
 9.7674 & 18.935 & 0.030 & R & FLWO \\ 
 9.8631 & 18.942 & 0.040 & R & FLWO \\ 
10.6708 & 19.097 & 0.040 & R & FLWO \\ 
10.8258 & 19.050 & 0.030 & R & FLWO \\ 
11.6529 & 19.246 & 0.030 & R & FLWO \\ 
11.6940 & 19.209 & 0.030 & R & FLWO \\ 
11.8158 & 19.258 & 0.030 & R & FLWO \\ 
12.6858 & 19.305 & 0.030 & R & FLWO \\ 
12.8158 & 19.358 & 0.040 & R & FLWO \\ 
13.6344 & 19.414 & 0.080 & R & FLWO \\ 
13.6590 & 19.480 & 0.080 & R & FLWO \\ 
19.6958 & 19.777 & 0.040 & R & FLWO \\ 
19.7354 & 19.807 & 0.040 & R & FLWO \\ 
19.7822 & 19.815 & 0.060 & R & FLWO \\ 
20.7058 & 19.859 & 0.040 & R & FLWO \\ 
20.8038 & 19.937 & 0.050 & R & FLWO \\ 
20.8398 & 19.795 & 0.050 & R & FLWO \\ 
21.6588 & 19.938 & 0.030 & R & FLWO \\ 
21.7862 & 19.949 & 0.040 & R & FLWO \\ 
21.8729 & 19.866 & 0.060 & R & FLWO \\ 
22.6098 & 19.945 & 0.050 & R & LCO40 \\ 
22.6448 & 20.017 & 0.040 & R & FLWO \\ 
22.7344 & 20.006 & 0.040 & R & FLWO \\ 
23.6582 & 20.080 & 0.030 & R & FLWO \\ 
23.7544 & 20.043 & 0.040 & R & FLWO \\ 
24.6514 & 20.146 & 0.040 & R & FLWO \\ 
24.8058 & 20.117 & 0.030 & R & FLWO \\ 
25.6658 & 20.158 & 0.020 & R & KPNO4m \\ 
25.7001 & 20.142 & 0.040 & R & FLWO \\ 
25.7098 & 20.148 & 0.020 & R & KPNO4m \\ 
25.7178 & 20.175 & 0.040 & R & FLWO \\ 
26.6718 & 20.255 & 0.040 & R & FLWO \\ 
27.7108 & 20.320 & 0.050 & R & FLWO \\ 
28.7200 & 20.229 & 0.050 & R & FLWO \\ 
28.7314 & 20.347 & 0.060 & R & FLWO \\ 
29.6629 & 20.422 & 0.040 & R & FLWO \\ 
29.6781 & 20.350 & 0.040 & R & FLWO \\ 
29.6895 & 20.422 & 0.050 & R & FLWO \\ 
30.6941 & 20.478 & 0.040 & R & FLWO \\ 
30.7201 & 20.493 & 0.040 & R & FLWO \\ 
31.7105 & 20.568 & 0.040 & R & FLWO \\ 
32.5908 & 20.549 & 0.070 & R & LCO40 \\ 
32.6581 & 20.588 & 0.040 & R & FLWO \\ 
33.6338 & 20.665 & 0.030 & R & KPNO4m \\ 
33.8010 & 20.563 & 0.060 & R & FLWO \\ 
33.8173 & 20.657 & 0.060 & R & FLWO \\ 
34.5648 & 20.700 & 0.040 & R & KPNO4m \\ 
34.6919 & 20.812 & 0.070 & R & FLWO \\ 
34.7081 & 20.705 & 0.050 & R & FLWO \\ 
36.8058 & 20.875 & 0.080 & R & FLWO \\ 
37.6631 & 20.820 & 0.070 & R & FLWO \\ 
37.6738 & 20.897 & 0.050 & R & FLWO \\ 
37.6846 & 20.823 & 0.060 & R & FLWO \\ 
38.7300 & 21.024 & 0.060 & R & FLWO \\ 
39.6716 & 21.012 & 0.070 & R & FLWO \\ 
51.7458 & 21.553 & 0.060 & R & FLWO \\ 
52.7058 & 21.698 & 0.110 & R & FLWO \\ 
53.7408 & 21.430 & 0.080 & R & FLWO \\ 
54.6944 & 21.309 & 0.090 & R & FLWO \\ 
57.7163 & 21.612 & 0.060 & R & FLWO \\ 
59.7069 & 21.456 & 0.070 & R & FLWO \\ 
60.7058 & 21.618 & 0.090 & R & FLWO \\ 
 0.6470 & 14.820 & 0.010 & I & FLWO \\ 
 0.6577 & 14.845 & 0.010 & I & FLWO \\ 
 0.6801 & 14.896 & 0.010 & I & FLWO \\ 
 0.6959 & 14.937 & 0.010 & I & FLWO \\ 
 0.6961 & 14.974 & 0.030 & I & KAIT \\ 
 0.7062 & 14.981 & 0.010 & I & FLWO \\ 
 0.7155 & 14.995 & 0.030 & I & KAIT \\ 
 0.7162 & 15.045 & 0.010 & I & FLWO \\ 
 0.7262 & 15.053 & 0.010 & I & FLWO \\ 
 0.7313 & 15.034 & 0.020 & I & KAIT \\ 
 0.7362 & 15.042 & 0.010 & I & FLWO \\ 
 0.7477 & 15.103 & 0.010 & I & FLWO \\ 
 0.7580 & 15.134 & 0.010 & I & FLWO \\ 
 0.7684 & 15.173 & 0.010 & I & FLWO \\ 
 0.7705 & 15.133 & 0.030 & I & KAIT \\ 
 0.7788 & 15.182 & 0.010 & I & FLWO \\ 
 0.7862 & 15.219 & 0.030 & I & KAIT \\ 
 0.7891 & 15.225 & 0.010 & I & FLWO \\ 
 0.7995 & 15.218 & 0.020 & I & FLWO \\ 
 0.8098 & 15.232 & 0.020 & I & FLWO \\ 
 0.8203 & 15.287 & 0.010 & I & FLWO \\ 
 0.8259 & 15.286 & 0.020 & I & KAIT \\ 
 0.8306 & 15.331 & 0.010 & I & FLWO \\ 
 0.8409 & 15.352 & 0.020 & I & FLWO \\ 
 0.8440 & 15.300 & 0.030 & I & KAIT \\ 
 0.8513 & 15.380 & 0.010 & I & FLWO \\ 
 0.8616 & 15.420 & 0.020 & I & FLWO \\ 
 0.8652 & 15.402 & 0.020 & I & KAIT \\ 
 0.8721 & 15.411 & 0.020 & I & FLWO \\ 
 0.8810 & 15.424 & 0.020 & I & KAIT \\ 
 0.8824 & 15.407 & 0.020 & I & FLWO \\ 
 0.8927 & 15.424 & 0.020 & I & FLWO \\ 
 0.8969 & 15.471 & 0.020 & I & KAIT \\ 
 0.9031 & 15.443 & 0.020 & I & FLWO \\ 
 0.9136 & 15.511 & 0.020 & I & FLWO \\ 
 0.9155 & 15.534 & 0.020 & I & KAIT \\ 
 0.9239 & 15.606 & 0.020 & I & FLWO \\ 
 0.9316 & 15.571 & 0.040 & I & KAIT \\ 
 0.9343 & 15.648 & 0.020 & I & FLWO \\ 
 0.9446 & 15.648 & 0.020 & I & FLWO \\ 
 0.9477 & 15.569 & 0.030 & I & KAIT \\ 
 0.9550 & 15.649 & 0.030 & I & FLWO \\ 
 1.6542 & 15.828 & 0.010 & I & FLWO \\ 
 1.7495 & 15.950 & 0.010 & I & FLWO \\ 
 1.8029 & 16.044 & 0.040 & I & KAIT \\ 
 1.8481 & 16.087 & 0.010 & I & FLWO \\ 
 1.8652 & 16.059 & 0.070 & I & KAIT \\ 
 1.9239 & 16.123 & 0.010 & I & FLWO \\ 
 2.8581 & 16.594 & 0.010 & I & FLWO \\ 
 2.9692 & 16.681 & 0.020 & I & FLWO \\ 
 3.7780 & 16.957 & 0.020 & I & FLWO \\ 
 3.7970 & 16.979 & 0.020 & I & FLWO \\ 
 3.8220 & 16.971 & 0.020 & I & FLWO \\ 
 3.8412 & 17.014 & 0.020 & I & FLWO \\ 
 3.8717 & 17.027 & 0.050 & I & FLWO \\ 
 4.6532 & 17.363 & 0.020 & I & FLWO \\ 
 5.6451 & 17.319 & 0.020 & I & FLWO \\ 
 5.7205 & 17.369 & 0.020 & I & LCO100 \\ 
 5.7224 & 17.365 & 0.020 & I & LCO100 \\ 
 6.5936 & 17.830 & 0.020 & I & LCO100 \\ 
 6.5956 & 17.878 & 0.030 & I & LCO100 \\ 
 6.6606 & 17.899 & 0.030 & I & FLWO \\ 
 6.8811 & 17.968 & 0.030 & I & FLWO \\ 
 7.6504 & 18.315 & 0.040 & I & FLWO \\ 
 7.8862 & 18.386 & 0.050 & I & FLWO \\ 
 8.6475 & 18.473 & 0.040 & I & FLWO \\ 
 8.7844 & 18.469 & 0.040 & I & FLWO \\ 
 8.8866 & 18.518 & 0.050 & I & FLWO \\ 
 9.6444 & 18.595 & 0.050 & I & FLWO \\ 
 9.7709 & 18.527 & 0.040 & I & FLWO \\ 
 9.8666 & 18.583 & 0.060 & I & FLWO \\ 
10.6807 & 18.741 & 0.050 & I & FLWO \\ 
10.8158 & 18.768 & 0.040 & I & FLWO \\ 
11.6667 & 18.906 & 0.040 & I & FLWO \\ 
11.7475 & 18.938 & 0.050 & I & FLWO \\ 
11.8441 & 18.985 & 0.060 & I & FLWO \\ 
 3.6058 & 15.935 & 0.020 & J & LCO40 \\ 
 4.6488 & 16.449 & 0.030 & J & LCO40 \\ 
 5.6552 & 16.416 & 0.020 & J & LCO40 \\ 
 6.6239 & 17.051 & 0.040 & J & LCO40 \\ 
 7.6887 & 17.438 & 0.030 & J & LCO40 \\ 
 8.6377 & 17.559 & 0.040 & J & LCO40 \\ 
 9.6427 & 17.736 & 0.040 & J & LCO40 \\ 
11.6355 & 18.233 & 0.070 & J & LCO40 \\ 
 3.6325 & 15.222 & 0.020 & H & LCO40 \\ 
 4.6757 & 15.575 & 0.030 & H & LCO40 \\ 
 5.6775 & 15.687 & 0.030 & H & LCO40 \\ 
 6.6493 & 16.199 & 0.040 & H & LCO40 \\ 
 7.6682 & 16.634 & 0.050 & H & LCO40 \\ 
 8.6645 & 16.833 & 0.080 & H & LCO40 \\ 
 9.6795 & 17.152 & 0.050 & H & LCO40 \\ 
\enddata
\tablecomments{[The complete version of this table is in the
electronic edition of the Journal.  The printed edition contains only
a sample.]}
\tablenotetext{a}{Days after 2003 March 29.4842 UT.}
\tablenotetext{b}{
FLWO: F.~L. Whipple Observatory 1.2-m telescope;
KAIT: 0.76-m Katzman Automatic Imaging Telescope;
LCO100: Las Campanas Observatory 2.5-m telescope (du~Pont);
KPNO4m: Kitt Peak National Observatory 4-m telescope;
Clay: Magellan Clay telescope;
LCO40: Las Campanas Observatory 1-m telescope (Swope)
}
\end{deluxetable}

\thispagestyle{empty}

\begin{deluxetable}{lccccccccccccl}
\tabletypesize{\small}
\rotate
\tablewidth{0pt}
\tablecaption{JOURNAL OF SPECTROSCOPIC OBSERVATIONS\label{specjournal}}
\tablehead{\colhead{$\Delta$T\tablenotemark{a}} & 
\colhead{UT Date\tablenotemark{b}} &
\colhead{Julian Day\tablenotemark{b}} &
\colhead{Tel.\tablenotemark{c}} &
\colhead{Range\tablenotemark{d}}  &
\colhead{Res.\tablenotemark{e}} &
\colhead{P.A.\tablenotemark{f}} &
\colhead{Par.\tablenotemark{g}} &
\colhead{Air.\tablenotemark{h}} & 
\colhead{Flux Std.\tablenotemark{i}} &
\colhead{See.\tablenotemark{j}} &
\colhead{Slit} &
\colhead{Exp.}  \\
\colhead{} &
\colhead{} &
\colhead{} &
\colhead{} &
\colhead{(\AA)} &
\colhead{(\AA)} &
\colhead{($^\circ$)} &
\colhead{($^\circ$)} &
\colhead{} &
\colhead{} &
\colhead{($^{\prime\prime}$)} &
\colhead{($^{\prime\prime}$)} &
\colhead{(s)} }
\startdata
0.72 & 2003-03-30.20 & 2452728.70 & FLWO & 3720-7540 & 6.4 & 0.0 &  -40.2  & 1.03 & F34         &     2    & 3.0 & 2$\times$1200 \\
0.75 & 2003-03-30.23 & 2452728.73 & MMT  & 3600-8700 & 8.0 & -16 &   -4.9 & 1.02 & CygOB2/H600 &   4.5     & 1.25 & 4$\times$900 \\
1.73 & 2003-03-31.21 & 2452729.71 & MMT  & 3600-8700 & 8.0 & -43 &  -26.8 & 1.03 & CygOB2/H600 &   1.5     & 1.25 & 3$\times$600   \\
2.66 & 2003-04-01.14 & 2452730.64 & MMT  & 3450-8650 & 6.4 & -65 &  -54.1 & 1.18 & F34/H600    &     1.5  & 1.0 &  900  \\
2.75 & 2003-04-01.23 & 2452730.73 & FLWO & 3720-7540 & 6.4 & 5.0 &  7.9   & 1.02 & F34         &     2    & 3.0 & 1200  \\
3.70 & 2003-04-02.18 & 2452731.68 & MMT  & 3450-8650 & 6.4 & -52 &  -40.6 & 1.06 & F34/HD84    &    2.5    & 1.0 &  4$\times$600   \\
3.82 & 2003-04-02.30 & 2452731.80 & FLWO & 3720-7540 & 6.4 & 53  &  62.2  & 1.13 & F34         &     2    & 3.0 & 2$\times$1800 \\
4.66 & 2003-04-03.14 & 2452732.64 & MMT  & 3450-8600 & 6.4 & -64 &  -52.5 & 1.15 & F34/HD84    &    2.5   & 1.0 &  900 \\
5.72 & 2003-04-04.22 & 2452733.72 & D25  & 3800-9320 & 7.7 & A.P.{\tablenotemark{k}}    &  -40   & 2.16 & L3218/L7379/L7987  & \nodata & 1.6    & 2$\times$600  \\
5.80 & 2003-04-04.28 & 2452733.78 & MMT  & 3450-8600 & 8.0 & 52 &   41.6 & 1.06 & F34/HD84    &    2    & 1.5 &  900    \\
6.60 & 2003-04-05.10 & 2452734.60 & D25  & 3800-9320 & 7.7 & A.P.{\tablenotemark{k}}     &  -9    & 1.59 & L7379/L7987 &  \nodata  & 1.6   & 900  \\
6.66 & 2003-04-05.14 & 2452734.64 & MMT  & 3500-8650 & 6.4 & -62 &  -61.0 & 1.11 & F34/HD84    &    2   & 1.0 &  2$\times$900  \\
7.67 & 2003-04-06.15 & 2452735.65 & MMT  & 3500-8650 & 6.4 & -61 &  -45.6 & 1.08 & F34/HD84    &    2.5    & 1.0 & 2$\times$1200  \\
8.78 & 2003-04-07.26 & 2452736.76 & MMT  & 3300-8450 & 8.0 & 48 &   50.8 & 1.05 & F34/HD84    &    3    & 1.25 &  3$\times$900  \\
9.63 & 2003-04-08.13 & 2452737.63 & D25  & 3800-9320 & 7.7 &  A.P.{\tablenotemark{k}}    &  6    & 1.58 &  L7379/L7987  &  \nodata & 1.6   & 2$\times$900 \\
9.67 & 2003-04-08.15 & 2452737.65 & MMT  & 3250-8300 & 6.4 & -61 &  -57.2 & 1.07 & F34/HD84    &    3   & 1.0 &  3$\times$900  \\
9.86 & 2003-04-08.34 & 2452737.84 & L3   & 3176-10400& 6-15& 188 & 54 &  1.21 &   F34/HD84   &  3.5  &  2  & 4$\times$2400 \\
10.68 & 2003-04-09.16 & 2452738.66 & MMT & 3200-8350 & 8.0 & -60 &  -45.0 & 1.08 & F34/HD84    &    3.5    & 1.25 & 3$\times$1200   \\
11.66 & 2003-04-10.16 & 2452739.66 & Clay& 3600-9000 & 10.4&  A.P.{\tablenotemark{k}}    &   -28     & 1.65 & F67/L3218    &  \nodata & 0.7 & 2$\times$900  \\
11.66 & 2003-04-10.14 & 2452739.64 & MMT & 3200-8350 & 8.0 & -62 &  -49.2 & 1.11 & F34/HD84    &    3.0    & 1.25 &  3$\times$900   \\
25.71 & 2003-04-24.19 & 2452753.69 & MMT & 3700-8100 & 6.4 & 10 &   13.6 & 1.02 & F34/HD84    &    1.5    & 1.0 & 3$\times$1800   \\
25.89 & 2003-04-24.37 & 2452753.87 & KII & 4110-9154 & 2.0 & -99 &   88.62 & 1.1 & F34              &    1.4    & 1.25 &  3$\times$600   \\
33.57 & 2003-05-02.05 & 2452761.55 &  Baade & 3400-8002 & 10.5 & 182 &  -11.5 & 1.61 & F66/HD84  &  1.2      & 1.0 & 4$\times$1800   \\
35.53 & 2003-05-04.01 & 2452763.51 &  Baade & 3400-8002 & 10.5 & 178 &    2.0 & 1.57 & F66/HD84  &   1.5     & 1.0 & 4$\times$1800   \\
36.55 & 2003-05-05.03 & 2452764.53 &  Baade & 6240-9220 & 5.1 & 179 &    180 & 1.57 & HD84  &   1.5     & 1.0 & 4$\times$1800   \\
55.90 & 2003-05-24.38 & 2452783.38 & KI  & 3260-9280 & 8.0 & 79.9 &  79.4 &  1.80 &  BD28/BD17   &    1.3      & 1.5    & 3$\times$900        \\
\enddata
\tablecomments{Not all spectra listed in this Table are shown in the
paper.}
\tablenotetext{a}{Days since 2003 March 29.4842 UT.}
\tablenotetext{b}{Midpoint of observation(s).}
\tablenotetext{c}{MMT = MMT 6.5m/Blue Channel; FLWO = FLWO 1.5m/FAST;
D25 = du~Pont 2.5m/WFCCD; L3 = Lick 3m/Kast; Clay = Magellan 6.5m
Clay/LDSS2; KII = Keck II 10m/ESI; Baade = Magellan 6.5m Baade/Boller
\& Chivens; KI = Keck I 10m/LRIS.}
\tablenotetext{d}{Observed wavelength range of spectrum.  In some cases, the
extreme ends are noisy, and are not shown in the figures.}
\tablenotetext{e}{Approximate spectral resolution (full width at half
maximum intensity), typically estimated from night-sky emission
lines.}
\tablenotetext{f}{Approximate average position angle of the
spectrograph slit.}
\tablenotetext{g}{Optimal parallactic angle over the course of the
exposures.}
\tablenotetext{h}{Average airmass of observations.}
\tablenotetext{i}{The standard stars are as follows: 
BD28 = BD+28$^{\circ}$4211, F34 = Feige~34, H600 = Hiltner~600---Stone (1977),
 Massey \& Gronwall (1990);
F66 = Feige~66, F67 = Feige~67, CygOB2 = Cyg~OB2~ No. 9---Massey et al. (1988);
HD84 = HD~84937, BD17 = BD+47$^{\circ}$4708---Oke \& Gunn (1983);
L3218 = LTT~3218, L7379 = LTT~7379, L7987 = LTT 7987---Hamuy et al. (1992; 1994).}
\tablenotetext{j}{Approximate seeing, estimated from the data and observers' records.}
\tablenotetext{k}{A.P. = At Parallactic.}
\end{deluxetable}

\begin{deluxetable}{ccccccc} 
\rotate
\tablecolumns{7} 
\tablewidth{0pc} 
\tablecaption{EMISSION LINE RATIOS\label{ratiotable}} 
\tabletypesize{\scriptsize} 
\tablehead{ 
\colhead{log } &  \colhead{log } & \colhead{log } &  
\colhead{log}  &  \colhead{} & \colhead{} & 
 \colhead{log}\\

\colhead{([O~III] $\lambda$5007/[O~II] $\lambda$3727)} &
 \colhead{([N~II] $\lambda$6584/[O~II] $\lambda$3727)} &
 \colhead{([O~II]+[O~III])/H$\beta $ } &
 \colhead{([N~II] $\lambda$6584/H$\alpha$) } & \colhead{H$\alpha $/H$\beta $ } &
 \colhead{H$\gamma$/H$\beta $ } & \colhead{([N~II] $\lambda$6584/[O~III] $\lambda$5007)} } \startdata
 0.2 & $<-0.9$ & 0.9 & $<-1.0$ & 3.0 & 0.4 & $ <-1.3$ \cr \enddata
\end{deluxetable}

\begin{deluxetable}{rlrcrrr}
\tabletypesize{\small}
\rotate
\tablewidth{0pt}
\tablecaption{GRB/SN FITS\label{fittable}}
\tablehead{\colhead{$\Delta$T\tablenotemark{a}} &
\colhead{Best SN\tablenotemark{b}} &
\colhead{SN \%\tablenotemark{c}} &
\colhead{Fit Range (\AA)\tablenotemark{d}} &
\colhead{Fit Error\tablenotemark{e}} &
\colhead{SN \% $B$\tablenotemark{f}} &
\colhead{SN \% $R$\tablenotemark{f}}}
\startdata
0.75  & SN~1997ef +44 &  5 & 4242-8000 & 0.025 &  2 &  6 \\
1.73  & SN~1997ef +44 &  5 & 4242-8000 & 0.025 &  2 &  6 \\
2.66  & SN~1997ef +84 &  2 & 4348-8000 & 0.087 &  1 &  3 \\
3.70  & SN~2002ap +28 &  2 & 4348-8000 & 0.570 &  1 &  2 \\
4.66  & SN~1997ef +25 &  6 & 4236-8000 & 0.134 &  2 &  7 \\
5.80  & SN~1997ef -11 &  1 & 4348-8000 & 0.056 &  1 &  1 \\
6.66  & SN~1998bw +11 &  8 & 3950-8000 & 0.110 &  4 & 10 \\
7.67  & SN~2002ap +6 &  6 & 4348-8000 & 0.853 &  3 &  8 \\
8.78  & SN~1998bw -7 & 13 & 3906-8000 & 0.459 & 11 & 15 \\
9.67  & SN~1998bw -6 & 26 & 3920-8000 & 0.135 & 21 & 30 \\
10.68 & SN~1997ef -11 & 19 & 4348-8000 & 2.949 & 14 & 22 \\
11.66 & SN~1998bw -3 & 38 & 4092-8000 & 1.067 & 29 & 43 \\
25.80 & SN~1998bw +6 & 61 & 3950-8000 & 0.695 & 46 & 68 \\
33.57 & SN~1998bw +13 & 63 & 3622-8000 & 2.154 & 45 & 72 \\  
35.53 & SN~1998bw +11 & 86 & 3950-8000 & 3.921 & 76 & 89 \\  
	   
\enddata
\tablenotetext{a}{Days since 2003 March 29.4842 UT.}

\tablenotetext{b}{SN spectrum that best matches the OT afterglow at
this epoch (SN and approximate phase relative to $B$-band maximum).
Note that the low SN fraction at early times ($\Delta T = 0.75-7.67$
days) makes the SN spectrum listed almost arbitrary.}
\tablenotetext{c}{Percentage of SN component in best-fit spectrum over
the fitting range listed.}  \tablenotetext{d}{Spectral range of the
overall fit.}  \tablenotetext{e}{Least-squares deviation from fit (not
a formal $\chi^2$ statistic).  Only the relative size of this number
is important.}  \tablenotetext{f}{Relative contribution of the SN in
the broad-band filter indicated, synthesized from the best-fit, scaled
SN spectrum.}
\end{deluxetable}

\end{document}